\documentclass[twocolumn,showpacs,twoside,10pt,prl,superscriptaddress]{revtex4-1}
\usepackage{graphicx}
\usepackage{epsfig}
\usepackage{epsf}
\usepackage{amssymb}
\usepackage{amsmath}
\usepackage{amsthm}
\usepackage{multirow}
\usepackage{mathrsfs}

\begin{document}

\title{ Dynamically Polarizing Spin Register of NV Centers in Diamond using Chopped Laser Pulses}

\author{Nanyang Xu}
\email{nyxu@hfut.edu.cn}
\affiliation{School of Electronic Science and Applied Physics, Hefei University of Technology, Hefei,  Anhui, 230009, China}

\author{Yu Tian}
\affiliation{School of Electronic Science and Applied Physics, Hefei University of Technology, Hefei,  Anhui, 230009, China}
\author{Bing Chen}
\affiliation{School of Electronic Science and Applied Physics, Hefei University of Technology, Hefei,  Anhui, 230009, China}
\author{Jianpei Geng}
\affiliation{School of Electronic Science and Applied Physics, Hefei University of Technology, Hefei,  Anhui, 230009, China}
\author{Xiaoxiong He}
\affiliation{School of Electronic Science and Applied Physics, Hefei University of Technology, Hefei,  Anhui, 230009, China}

\author{Ya Wang}
\email{ywustc@ustc.edu.cn}
\affiliation{Hefei National Laboratory for Physical Sciences at the Microscale and Department of Modern Physics,
University of Science and Technology of China, Hefei 230026, China}
\affiliation{CAS Key Laboratory of Microscale Magnetic Resonance, USTC}
\affiliation{Synergetic Innovation Center of Quantum Information and Quantum Physics, USTC}

\author{Jiangfeng Du}
\email{djf@ustc.edu.cn}
\affiliation{Hefei National Laboratory for Physical Sciences at the Microscale and Department of Modern Physics,
University of Science and Technology of China, Hefei 230026, China}
\affiliation{CAS Key Laboratory of Microscale Magnetic Resonance, USTC}
\affiliation{Synergetic Innovation Center of Quantum Information and Quantum Physics, USTC}

\begin{abstract}
Nuclear spins nearby nitrogen-vacancy (NV) centers in diamond are excellent quantum memory for quantum computing and quantum sensing, but are difficult to be initialized due to their weak interactions with the environment. Here we propose and demonstrate a magnetic-field-independent, deterministic and highly efficient polarization scheme by introducing chopped laser pulses into the double-resonance initialization method. With this method, we demonstrate initialization of single-nuclear-spin approaching $98.1\%$ and a $^{14}N$-$^{13}C$ double-nuclear-spin system approaching $96.8\%$ at room temperature. The initialization is limited by a nuclear-spin depolarization effect due to chopped laser excitation. Our approach could be extended to NV systems with more nuclear spins and would be a useful tool in future applications such as nano-MRI and single-cell NMR.
\end{abstract}
\maketitle

\begin{figure}[t]
\centering
\includegraphics[width=1\columnwidth]{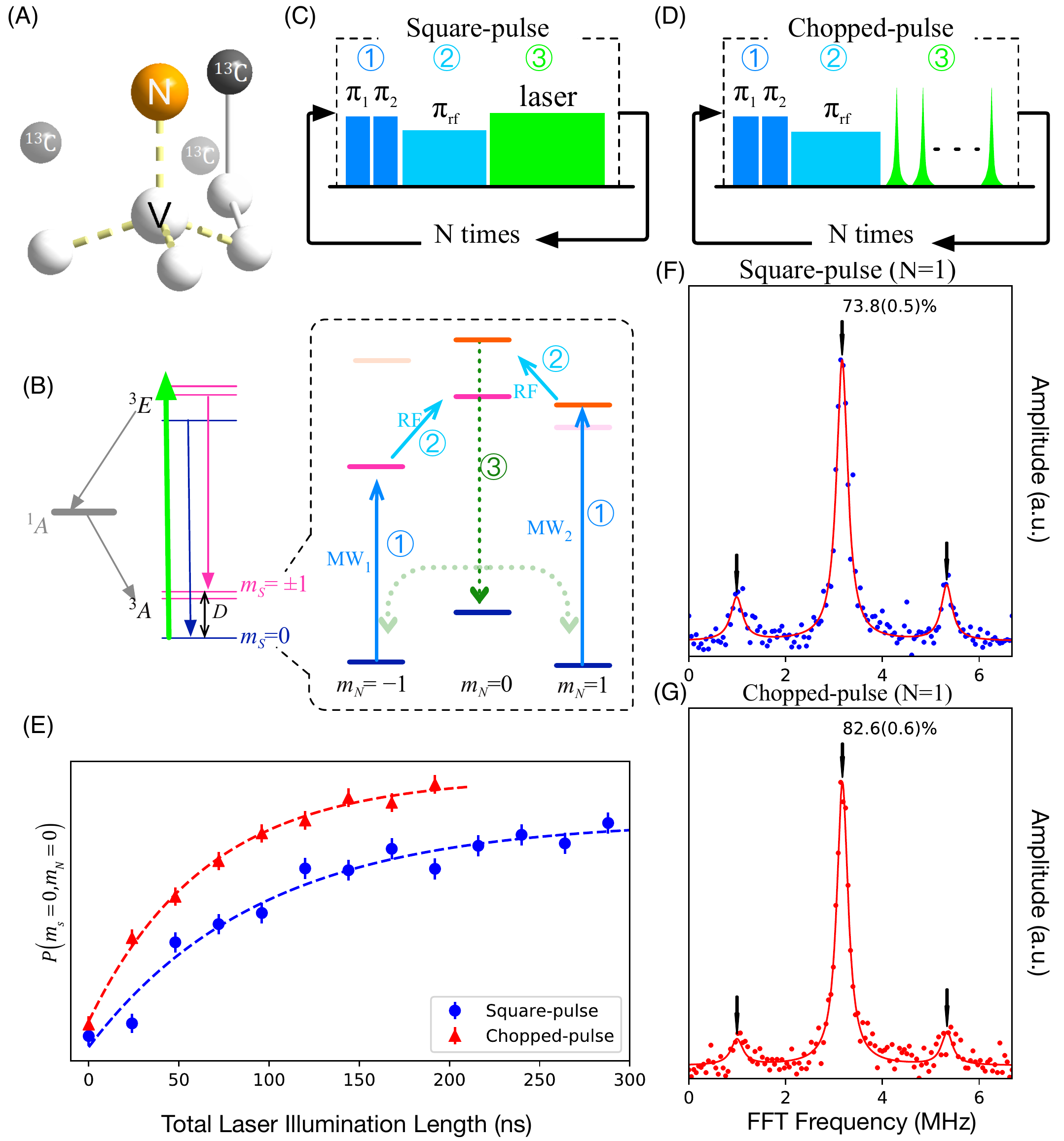}
\caption{(color online). Idea and experimental effect of chopped laser pulses in double-resonance DNP. A) schematics of NV center in diamond, with $^{13}C$ nuclear spins in nearby lattices. B) simplified energy-level diagram and transitions of NV electron states with an substitutional $^{14}N$ nuclear spin. The spin Hamiltonian is $H/2\pi=DS^2_z + \mu_eBS_z + QI^2_z + AS_zI_z-\mu_NBI_z $  where $D=2.87$GHz ($Q=-4.945$MHz) is the quadrupole splitting of the NV electron ($^{14}N$ nuclear) spin in the ground state, $\mu_e$($\mu_N$) is the electron(nuclear) gyromagnetic ratio  and $A= -2.16$MHz is the hyperfine interaction strength. C-D) pulse sequences for (C) original- and (D) chopped-pulse double-resonance DNP. E) probability on state $m_N=0$ and $m_s=0$ after population transfer to $^{14}N$ spin using both square and chopped laser pulses. The probability is calculated from the fluorescence measured after DNP with a varying-length laser illumination and two subsequent selective MW pulses in the $m_s=\pm1$ subspaces. F-G) FFT spectrum of the electron time-domain Ramsey signals after DNP using (F) square and (H) chopped laser pulses. The length of total pumping laser pulses is 300ns in (F) and 160ns in (G). The red line is Lorentzian fit of the data. The polarization is calculated from the fitted amplitudes while the error from the integration of noise in the adjacent area of the peak. The experiments are looped over $10^6$ times for signal accumulation. }
\label{fig1}
\end{figure}

\begin{figure}[t]
\centering
\includegraphics[width=1\columnwidth]{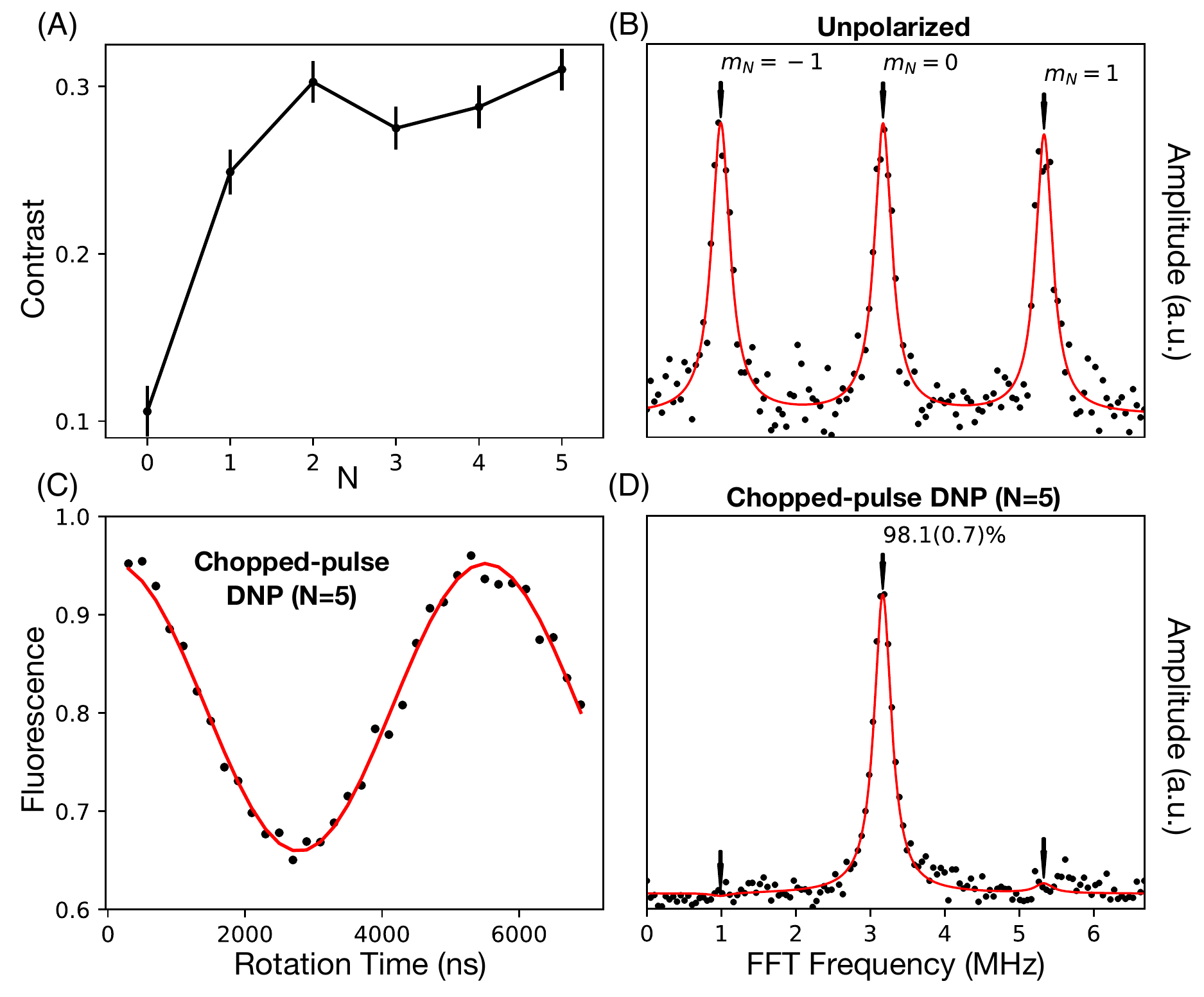}
\caption{(color online). Details of polarizing a $^{14}N$ nuclear spin. A) result of the recursive DNP experiment. The contrast is calculated from the fluorescence at $m_s=0$ state and $m_s=-1$ state in the nuclear-spin $m_N=0$ subspace after chopped-pulse DNP iterations.  B,D) FFT spectrum of the electron time-domain Ramsey signals measured with (B) no DNP and (D) $N=5$ chopped-pulse DNP iterations. The notation $m_N$ in the diagram denotes the transition in the manifold of $^{14}N$ nuclear spin states. The red line is Lorentzian fit of the data. The data in (D) is fitted with given positions from the unpolarized spectrum in (B). The polarization is calculated from the fitted amplitudes while the error from the integration of noise in the adjacent area of the peak. C) electron-spin Rabi oscillation after $N=5$ chopper-pulse DNP iterations in the $m_N=0$ subspace. The experiments are looped $5\times 10^5$ times in (C) and over $10^6$ times in (A,B,D) for signal accumulation.} \label{fig2}
\end{figure}

Nuclear spins in solid-state platforms such as diamond are important quantum resources because of their extremely-long coherence time. Besides being natural quantum memories\cite{lukin_memory, quant_memory_awschalom,single_shot_electron,rt_ent_2013,quant_entangle_2014,hanson_ent_measurement,ys_qrepeater,hanson_quantnetwork}, nuclear spins are developed as multi-spin quantum register implementing quantum algorithms\cite{lukin_science_2007,nv_3bit_entangle,nv_qec,hanson_qec,hanson_weak_nuclear_spin,kf_prl_2016,nv_aqc}, and ancilla qubit for enhanced quantum sensing\cite{Joreg_ent_sensing,Joerg_broadband,fador_qec_sensing,Joerg_sensing_cluster,Degen_npj_2017,Joerg_chemical_shift}. However, the tiny thermal nuclear spin polarization represents a major obstacle to reach their full potential. This limits nuclear spins often being partially used, for example, via state selection\cite{nv_3bit_entangle,nv_qec,hanson_qec,Joerg_sensing_cluster,Joerg_chemical_shift}. But the success rate decreases exponentially as the number of nuclear spins increases. Therefore, deterministic and efficient polarization of nuclear spins are highly desirable to fully explore the capability of nuclear spins.

Transferring polarization from electron spin to nuclear spins, \emph{i.e.}, the dynamical nuclear polarization (DNP), is a standard and deterministic strategy. For nitrogen-vacancy(NV) center in diamond, its optically polarized electron spin \cite{Joerg_science_1997,suter_npol} provides a good means to polarize the surrounding nuclear spins. Unlike bulk nuclear spin polarization for sensitive nuclear magnetic resonance imaging\cite{nc_frydman_2015,nc_pines_2015,nc_hollenberg_2018}, we focus on initializing a multi-nuclear-spin-qubit spin hybrid system by spin polarization technique as shown in Fig.\ref{fig1}A. It has recently been shown that more than 10 nuclear spins can be coherently detected around an NV electron spin\cite{prb_nuclear_spin, Taminiau_22_nuclspin}. The nuclear spins have been hyperpolarized using optical pumping\cite{nv_n_pol, Awschalom_polarization}. It utilizes level anti-crossing (LAC) of coupled electron-nuclear spin system, which occurs only at specific magnetic fields around 500 G (excited-state LAC) or 1000 G (ground-state LAC)\cite{nv_n_pol, bulk_c_pol2013, yangwen_scientic_report, ensem_c_pol2013,gjp_prl_2016, rx_nc_2015, duan_geometric_gates}.

Away from the anti-crossing the double-resonance method is developed and transfers the polarization through a combination of microwave(MW), radio-frequency(RF) and optical fields\cite{lukin_science_2007,single_shot_electron}(Fig.\ref{fig1}C). This method is independent of magnetic field and general in various applications. The selective $\pi$ pulses imposed on the electron and nuclear spin respectively, performs a SWAP-like operation by exchanging the polarization between the electron spin and the nuclear spin. The following square laser pulse then re-initializes the electron spin. However, it has been demonstrated that the nuclear spin depolarizes under optical illumination, especially under low magnetic fields\cite{single_shot_electron, Jiang_nuclear_coherence, single_shot_nuclear, cspin_corrspec}. Consequently, the double-resonance method still suffers from a low degree of nuclear spin polarization of $77\%$\cite{dr_n_pol}, even under recursive polarization\cite{dr_n_pol, wy_qsim_2015}.

In this letter, we propose and demonstrate an alternative double-resonance DNP method based on chopped laser pulses. With this method, we demonstrate initialization of single-nuclear-spin approaching $98.1\%$ and a $^{14}N$-$^{13}C$ double-nuclear-spin system approaching $96.8\%$ at room temperature. The initialization is limited by a nuclear-spin depolarization effect due to chopped laser excitation. Finally, we provide a model to understand how chopped laser technique helps and analyze the double-resonance DNP process. It show a good agreement between theoretical simulations and experimental results.

As shown in Fig.\ref{fig1}D, our idea is using very-shot (chopped) laser pulses to repeatedly pump the electron spin. Under laser excitation, the electron spin undergoes a spin-dependent cycling transition(the left panel in Fig.\ref{fig1}B). For $m_s=0$ state, it will be excited to the corresponding $m_s=0$ state in $^3E$ and then go back to original state. However, for $m_s=\pm 1$ state, it has high probability to jump to the intermediate state $^1A$ and go to the ground state $m_s=0$. Between the laser pulses, a finite idle wait time is imposed to taking into account the finite lifetime of the excited states. In such a way several rounds of excitation is enough to initialize the electron spin into $m_s=0$ state. More importantly, the NV center has less probability losing one electron, which is known as charge-state conversion \cite{Manson_nv0,Waldherr_nv0,Joerg_nv0,Siyushev_nv0,si_npol}, by using the chopped laser pulses than by using the square laser pulse. This is understood by noting that once being in the excited state, the electron has certain probability be further pumped to the conduction band during the rest of the square laser pulse. In the contrast, the chopped laser pulses let the electron relaxes back to ground state via applying the idle wait time in between and thus reduce this probability. This is shown later to be important for the observed efficient DNP at low magnetic field.

To realize our idea, we perform experiments with an NV center in a bulk diamond on a room-temperature optically detected magnetic resonance (ODMR) system. A green laser of 532 nm is applied to the sample and the fluorescence ranging from 650 to 800 nm is collected via the same confocal microscopy system. The laser is focused by a 100X oil-immersed objective with a spot size around 300nm in diameter. A static magnetic field $B$ ranging from 5 to 10 G is applied along the NV axis to split the $m_s=\pm1$ sub-levels. The microwave fields is generated from an IQ-modulation system where we use an arbitrary waveform generator (Tektronix  AWG520) to synthesize different frequencies and phases, and radiated to the sample via a coplanar waveguide.  To generate the chopped laser pulse, we use a strongly-focused acousto-optic modulator (AOM) with a bandwidth of 350MHz to modulate the laser beam. The AOM is driven by a high-speed AWG (Tektronix AWG610) which has an output bandwidth over 800 MHz at a sampling rate of 2.6GSPS maximally. In order to suppress the laser leakage, another AOM is used following the first one, which turns off the laser in the rest time of experiment. We generated pieces of laser pulse as short as 4$ns$ (AWG output) in experiment and used an idle wait time ranging from 20 to 30$ns$ because the DNP efficiency does not change obviously for a longer wait time up to 300$ns$ \cite{si_npol}.

We test our idea using an NV center (NV1) with its substitutional $^{14}N$ nuclear spin. The description of the electron-nuclear system is in Fig.\ref{fig1}B. The electron population in $m_N=\pm1$ subspaces is transferred to $m_N=0$ subspace using the double-resonance method. Then we apply a laser pulse with different lengths (or the number of pulses in chopped-pulse scheme). Fig.\ref{fig1}E shows the monitored fluorescence which represents the probability at the state $|m_N=0,m_s=0\rangle$. In both cases, we observe a gradually-growing curve which characterizes the re-initializing process of electron spin as the length of laser illumination increases. Interestingly, there is a clear probability difference in the flat area which marks the nuclear spin polarization difference after re-initializing the electron spin. To confirm this observation, we measure the corresponding electron time-domain Ramsey signals using the laser length in the flat area and show the FFT spectrums in Fig.\ref{fig1}F and Fig.\ref{fig1}G . The difference between polarizations in each case is nearly $9\%$.  Note that the illumination length does not include the idle time in between the chopped pulses in the whole manuscript.

The results of repetitive application of the chopped-pulse DNP are shown in Fig.\ref{fig2}. Fig.\ref{fig2}A displays the evolution of fluorescence contrast contributed by $m_N=0$ manifold as iteration increases. The contrast is calculated from $(I_{m_s=0,m_N=0}-I_{m_s=-1,m_N=0})/I_0$, where $I_{m_s=0,m_N=0}-I_{m_s=-1,m_N=0}$ is obtained by measuring the fluorescence intensity after a selective electron $\pi$ pulse and $I_0$ is obtained without applying the selective pulse. The contrast increases 3 times after five iterations, indicating that the nitrogen nuclear spin is pumped from an initial thermal state into an almost pure $m_N=0$ state. This is consistent with the FFT spectrum measurement as shown in Fig.\ref{fig2}B (N=0, thermal state) and Fig.\ref{fig2}C (N=5). By fitting the spectrum and comparing
amplitudes of the peaks at corresponding frequencies, we realize an initialization of $98.1(0.7)\%$. Fig.\ref{fig2}D shows the selective Rabi oscillation for $m_N=0$ state. The fitted contrast is $\sim30\%$ approaching the maximal contrast, which further confirms our observation.

\begin{figure}[t]
\centering
\includegraphics[width=1\columnwidth]{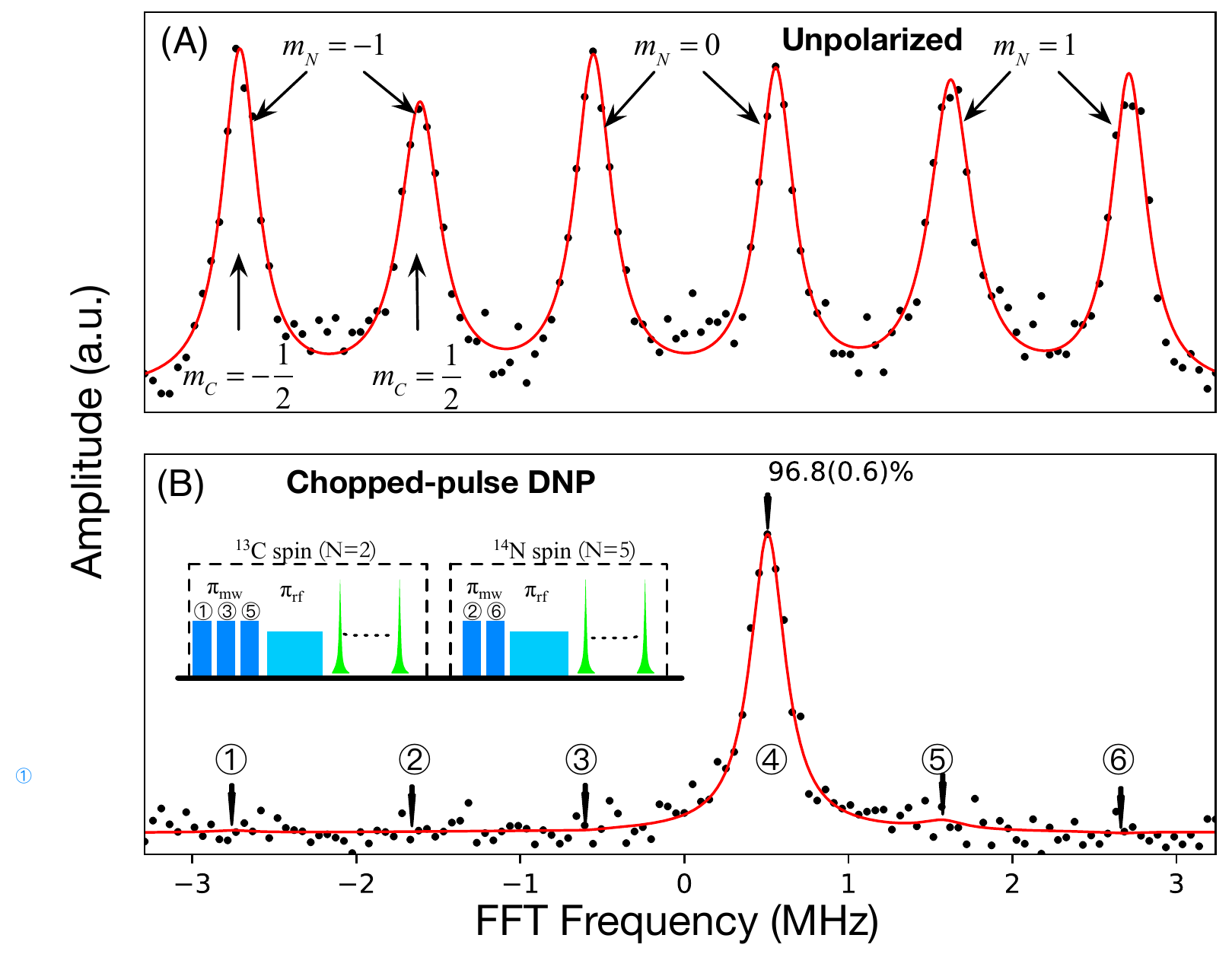}
\caption{(color online). Result of polarizing a $^{14}N$-$^{13}C$  double-nuclear-spin system. A-B) FFT spectrum of the quadrature electron time-domain Ramsey signals measured with (A) no DNP and (B) $N=(5,2)$ chopped-pulse DNP iterations for $^{14}N$ and $^{13}C$. The pulse sequence used in (B) is plotted inset. The electron-$^{13}C$ hyperfine interaction is around 1.1MHz. The notation $m_N$($m_C$) in the diagram denotes the transition in the manifold of $^{14}N$($^{13}C$) nuclear spin states.  The red line is Lorentzian fit of the data. The data in (B) is fitted with given positions from the unpolarized spectrum in (A).  The polarization is calculated from the fitted amplitudes while the error from the integration of noise in the adjacent area of the peak. The experiments are looped around $1.5\times 10^7$ times in (A) and $10^6$ times in (B) for signal accumulation. } \label{fig3}
\end{figure}

A highly-polarized multi-nuclear-spin register is an important quantum resource, but is still hard to be achieved. We further demonstrate that our method can be extended to more than one nuclear spin. Fig.\ref{fig3}A shows the FFT spectrum of a single NV center couples to a $^{13}C$ nuclear spin with the hyperfine splitting around 1.1MHz (NV2). The six peaks is associated to different nuclear spin state of $^{14}N$ and $^{13}C$. The DNP procedure is applied sequentially on these two nuclear spins. By applying chopped-pulse DNP $N=5$ times for $^{14}N$ and $N=2$ times for $^{13}C$ spin, Fig.\ref{fig3}B shows an efficient polarization of both nuclear spins with only one peak remains visible. From the data we realize an initialization of $96.8(0.6)\%$. This result means that we achieve a deterministic and efficient initialization of a three qubit quantum register by including the electron spin.

\begin{figure}[t]
\centering
\includegraphics[width=1\columnwidth]{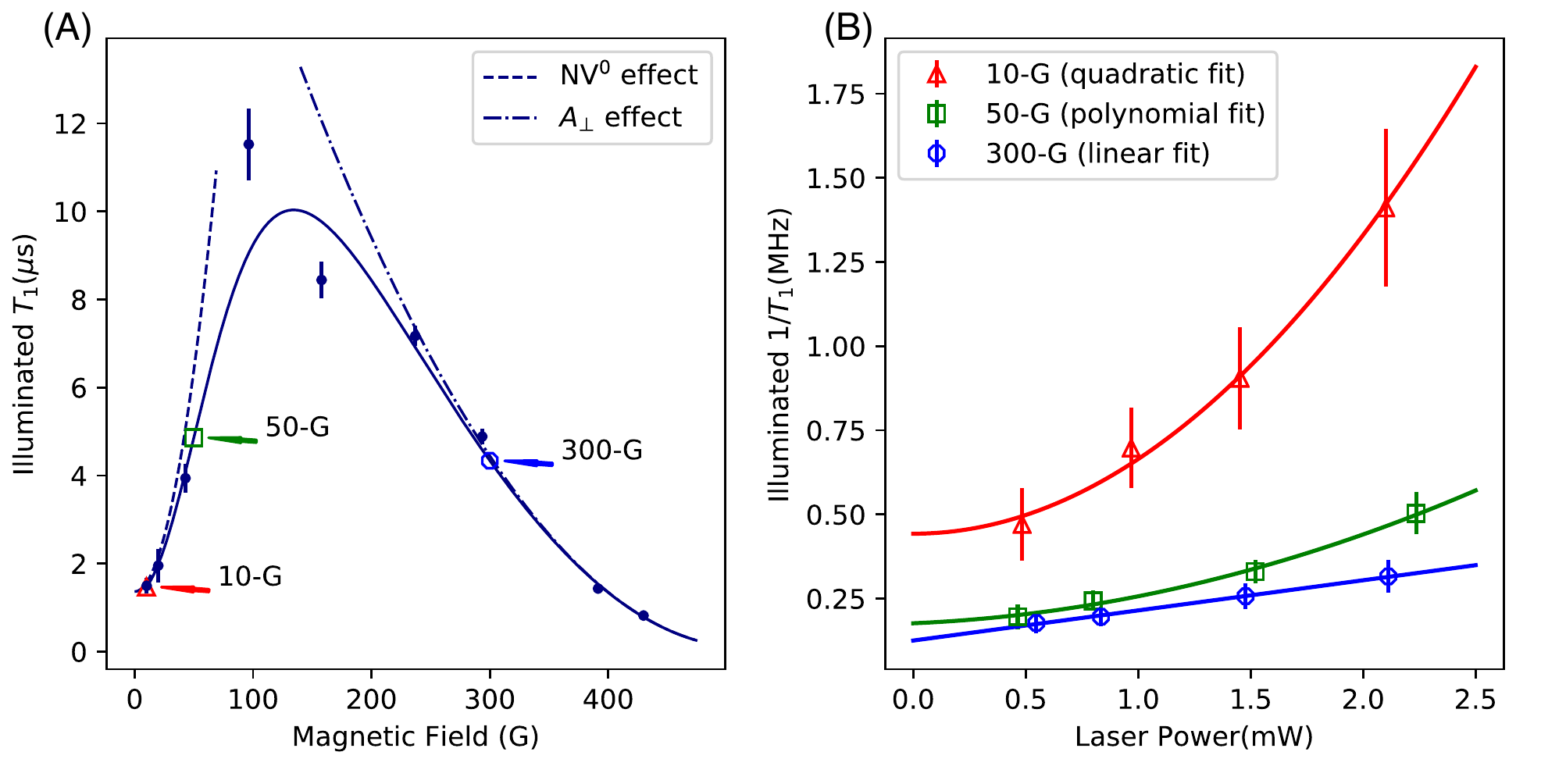}
\caption{(color online). Dependence of Illuminated nuclear depolarization on axial magnetic field and laser power. A) The measured illuminated nuclear $T_1$ (points) with different magnetic fields which are aligned along the NV axis. The solid line is a fit of the points using two independent quadratic function (dashed lines) considering the $A_{\bot}$ and NV$^{0}$ effect respectively. And the dashed line each shows an individual effect. B) the measured depolarization rates ($1/T_1$) with different laser powers. The 10-G (300-G) line in red (blue) is fit with a quadratic (linear) function. And the 50-G line (green) is fit with polynomial (up to quadratic) function.} \label{fig_mech_laser}
\end{figure}

\begin{figure}[t]
\centering
\includegraphics[width=1\columnwidth]{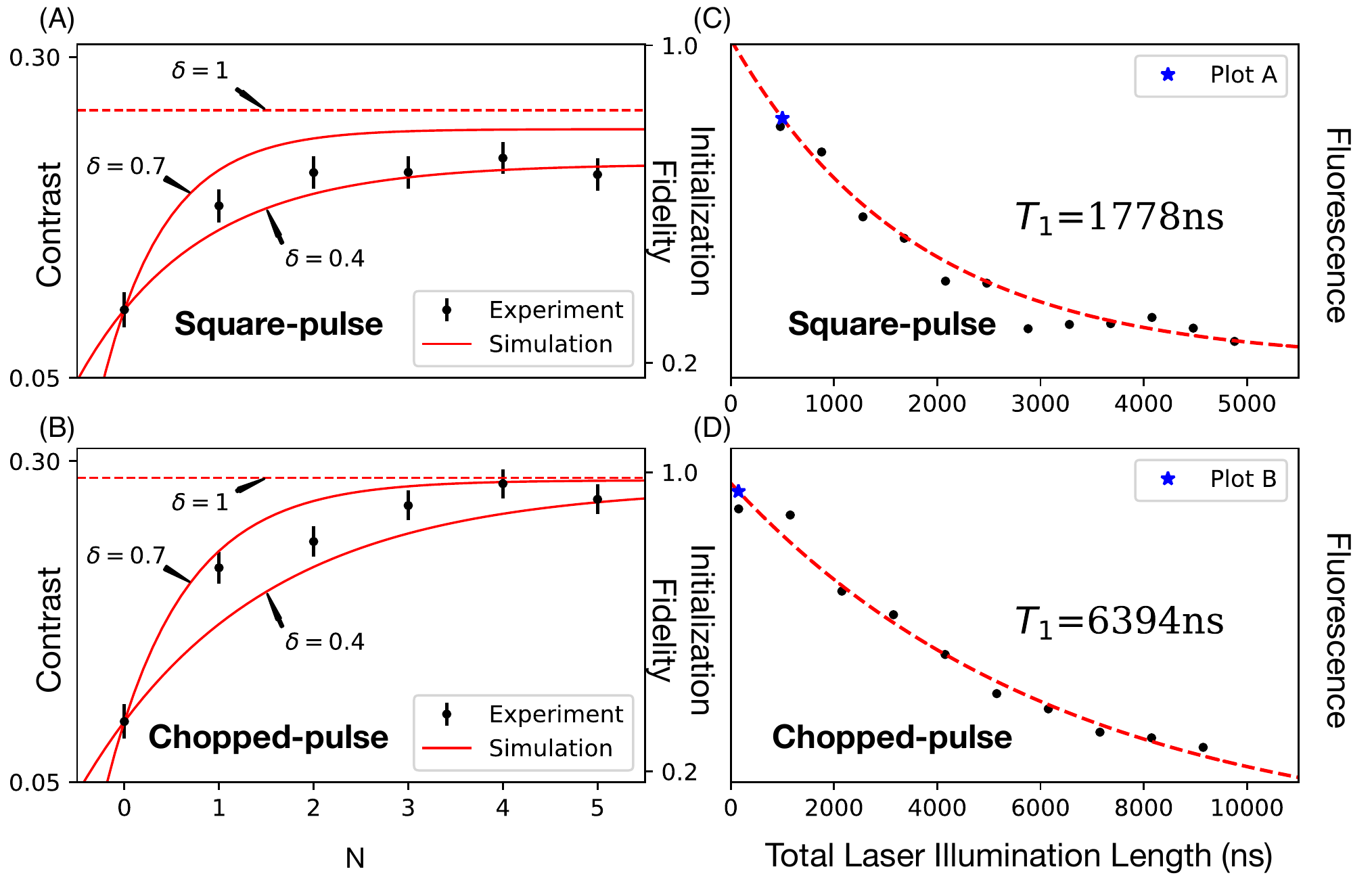}
\caption{(color online). Comparison between simulation results and experimental measurements. A-B) The experimental (points) and simulated (red line) contrast using (A) square-pulse and (B) chopped-pulse DNP iterations. The total illuminated length is 500$ns$ in (A) and 148$ns$ in (B) to fully re-polarize the electron spin. The red line is contrast calculated from the simulation while the points are measured data in experiment. The right y axis is  the expected initialization fidelity calculated from the contrast. C-D) The illuminated nuclear depolarization using (C) square and (D) chopped laser pulses. The red dashed line is an exponential fit of the points with $T_1$=1.8$\mu s$ in (C) and 6.4$\mu s$ in (D). This difference is possibly due to the actual laser amplitudes in the two cases, which is mainly limited by the bandwidth of the AOMs. The data is measured using the same experimental sequence as in Fig.\ref{fig1}E. The corresponding illumination length used in (A) and (B) are denoted as stars. The experiments are looped over $10^6$ times for signal accumulation.} \label{fig_mech_dnp}
\end{figure}

To gain deep insight into this process, we investigated the illuminated $T_1$ time of $^{14}N$ nuclear spin under different magnetic field (Fig.\ref{fig_mech_laser}A). In the field regime close to 500 G (the excited-state LAC point), we observed a quadratic-scaling $T_1$ time on magnetic fields where $1/T_1 \propto 2{A_{\bot}}^2/({2{A_{\bot}}^2+{\delta}^2})$, with $\delta = D-\mu_eB$ and $A_{\bot}\simeq 40$MHz ($D\simeq 1425$MHz) is the perpendicular component of hyperfine interaction \cite{single_shot_nuclear}.  This is due to the excited-state flip-flop process induced by hyperfine interaction, consistent with previous observations under fields above 500 G\cite{single_shot_nuclear}. However, in the low fields below 150 G, we observed a completely-different dependence which has not been reported before. To understand this, we measure the depolarization rate with different laser powers as shown in Fig.\ref{fig_mech_laser}B. Interestingly, a linear-scaling relation is observed in a field of 300 G while it is quadratic in 10 G. The former is easily understood, since the dwelling time in the excited state is linear with laser power. In the latter case, we attribute it to the charge-state conversion (NV$^{-}$ to NV$^{0}$ ), which shows the same quadratic dependence on laser power \cite{Joerg_nv0}. In this situation, the chopped pulse lowers the conversion rate from NV$^{-}$ to NV$^{0}$ and reduces the depolarization of nuclear spins (Fig.\ref{fig_mech_dnp}C-D). This new dependence also explains why the charge effect on the nuclear spin depolarization is not observed at high magnetic fields. From Fig.\ref{fig_mech_laser}A one can see this new scaling is also quadratic at low magnetic fields. It is probable due to the presence of an energy crossing point around zero Gauss for NV$^{0}$ state.

A toy model has been provided in \cite{dr_n_pol} to analyze the performance of the DNP iterations.
Here, we introduce a more-precise model to describe the process. Starting from electron spin state $m_s=0$ in Fig.\ref{fig1}B , there's initial populations on nuclear states $P_{0}(0)= c$ and $P_{\pm 1}(0) = 1-c$, where $P_i(j)$ denotes the population at state $m_N=i$ after $j$ iterations and $c=1/3$ for the $^{14}N$ spin. For the $j$-th iteration, there's $P_{\pm 1}(j-1)\delta$ population transferred from state $m_N=\pm 1$ to  state $m_N=0$, where $1>\delta > 0$ is the efficiency of the population transfer. And in the laser pumping process after population transfer, the population $p$ at state $m_N=0$ decays as $\mathscr{T}_1(p)=(p-c)e^{-\tau/T_1}+c$, where $\tau$ is the total laser-illumination length and $T_1$ the illuminated nuclear depolarization time. Then we can calculate the population at state $m_N=\pm 1$ after $j$ iterations is $P_{\pm 1}(j) = (1-c)(\alpha ^{j} + \frac{1-\alpha^j}{1-\alpha}\beta)$, where $\alpha = (1-\delta)e^{-\tau/T_1}$ and $\beta = 1-e^{-\tau/T_1}$. The polarization here equals to the population at state $m_N=0$, which is $P_0(j) = 1-P_{\pm 1}(j)$. In an ideal case where no nuclear depolarization happens, \emph{i.e.}, $e^{-\tau/T_1} = 1$, a perfect DNP is achieved with a polarization of 100\% as $j\to\infty$. In a more general case, the polarization approaches an upper limit $P_0(\infty) = 1-(1-c)( \frac{\beta}{1-\alpha})$ determined by the nuclear depolarization $e^{-\tau/T_1}$ and transfer efficiency $\delta$ together.

Fig.\ref{fig_mech_dnp} shows the simulation results according to this mechanism comparing with experimental measurements. The target process is to polarize the $^{14}N$ spin on NV2 using the square- or chopped-pulse schemes. We first measure the nuclear illuminated depolarization in Fig.\ref{fig_mech_dnp}C-D and calculate the $T_1$ by fitting the data. Then the contrasts are measured after the DNP iterations in experiments and shown as points in Fig.\ref{fig_mech_dnp}A-B. Note that, the secondary $y$ axis denotes the expected polarization. The lines are simulation results and the dashed one marks the upper limit $1-(1-c)\beta$ when $\delta = 1$. Most of the experimental points are located in the area with $\delta$ in a  range from 0.4 to 0.7. 

From this result, we can see a notable difference on the upper limit (the $\delta$=1 line) in Fig.\ref{fig_mech_dnp}A and Fig.\ref{fig_mech_dnp}B. This is due to the difference of nuclear depolarizations in the square- and chopped-pulse schemes. Another important difference is, the $\delta$=0.7 line is much closer to the $\delta$=1 line in Fig.\ref{fig_mech_dnp}B than Fig.\ref{fig_mech_dnp}A. This means that the upper limit is easier to be achieved in the weakly-depolarizing case than in the strongly one, since the former requires a lower $\delta$ than the latter. These two differences form the reason why the chopped-pulse DNP achieves a much better polarization than previous double-resonance methods.

To conclude, we propose and demonstrate an alternative double-resonance DNP scheme based on chopped laser pulses that can be applied in arbitrary magnetic field. We achieve a $^{14}N$ initialization of $98.1\%$ and a $^{14}N$-$^{13}C$ initialization of $96.8\%$ in a magnetic field below 10 G. The initialization is limited by a nuclear-spin depolarization effect due to charge state conversion at low magnetic field. This method is especially useful at low temperatures where the LAC-method cannot be applied. Current advances in quantum control technologies allow more $^{13}C$ nuclear spins to be detected and coherently controlled\cite{Taminiau_22_nuclspin}. These weakly-coupled nuclear spins have much longer illuminated nuclear spin relaxation time. Our method thus provide a good initialization method for all these nuclear spins which is important for their future application in quantum computing and quantum sensing.

The authors acknowledge financial support by the National Key R\&D Program of China (Grants No. 2018YFA0306600, 2017YFA0305000 and 2018YFF01012505 ), the NNSFC (Grant No. 61376128, 11775209, 11604069), the Fundamental Research Funds for the Central Universities, the Anhui Provincial Natural Science Foundation(Grant No. 1708085QA09), the Innovative Program of Development Foundation of Hefei Center for Physical Science and Technology (Grants No. 2017FXCX005), the CAS (Grants No. GJJSTD20170001 and QYZDY-SSW-SLH004), Anhui Initiative in Quantum Information Technologies (Grant No. AHY050000) and the Thousand Young Talents Plan. All the data diagrams are plotted with the graphical library Matplotlib\cite{matplotlib}.


\begin{thebibliography}{47}%
\makeatletter
\providecommand \@ifxundefined [1]{%
 \@ifx{#1\undefined}
}%
\providecommand \@ifnum [1]{%
 \ifnum #1\expandafter \@firstoftwo
 \else \expandafter \@secondoftwo
 \fi
}%
\providecommand \@ifx [1]{%
 \ifx #1\expandafter \@firstoftwo
 \else \expandafter \@secondoftwo
 \fi
}%
\providecommand \natexlab [1]{#1}%
\providecommand \enquote  [1]{``#1''}%
\providecommand \bibnamefont  [1]{#1}%
\providecommand \bibfnamefont [1]{#1}%
\providecommand \citenamefont [1]{#1}%
\providecommand \href@noop [0]{\@secondoftwo}%
\providecommand \href [0]{\begingroup \@sanitize@url \@href}%
\providecommand \@href[1]{\@@startlink{#1}\@@href}%
\providecommand \@@href[1]{\endgroup#1\@@endlink}%
\providecommand \@sanitize@url [0]{\catcode `\\12\catcode `\$12\catcode
  `\&12\catcode `\#12\catcode `\^12\catcode `\_12\catcode `\%12\relax}%
\providecommand \@@startlink[1]{}%
\providecommand \@@endlink[0]{}%
\providecommand \url  [0]{\begingroup\@sanitize@url \@url }%
\providecommand \@url [1]{\endgroup\@href {#1}{\urlprefix }}%
\providecommand \urlprefix  [0]{URL }%
\providecommand \Eprint [0]{\href }%
\providecommand \doibase [0]{http://dx.doi.org/}%
\providecommand \selectlanguage [0]{\@gobble}%
\providecommand \bibinfo  [0]{\@secondoftwo}%
\providecommand \bibfield  [0]{\@secondoftwo}%
\providecommand \translation [1]{[#1]}%
\providecommand \BibitemOpen [0]{}%
\providecommand \bibitemStop [0]{}%
\providecommand \bibitemNoStop [0]{.\EOS\space}%
\providecommand \EOS [0]{\spacefactor3000\relax}%
\providecommand \BibitemShut  [1]{\csname bibitem#1\endcsname}%
\let\auto@bib@innerbib\@empty
\bibitem [{\citenamefont {Maurer}\ \emph {et~al.}(2012)\citenamefont {Maurer},
  \citenamefont {Kucsko}, \citenamefont {Latta}, \citenamefont {Jiang},
  \citenamefont {Yao}, \citenamefont {Bennett}, \citenamefont {Pastawski},
  \citenamefont {Hunger}, \citenamefont {Chisholm}, \citenamefont {Markham},
  \citenamefont {Twitchen}, \citenamefont {Cirac},\ and\ \citenamefont
  {Lukin}}]{lukin_memory}%
  \BibitemOpen
  \bibfield  {author} {\bibinfo {author} {\bibfnamefont {P.~C.}\ \bibnamefont
  {Maurer}}, \bibinfo {author} {\bibfnamefont {G.}~\bibnamefont {Kucsko}},
  \bibinfo {author} {\bibfnamefont {C.}~\bibnamefont {Latta}}, \bibinfo
  {author} {\bibfnamefont {L.}~\bibnamefont {Jiang}}, \bibinfo {author}
  {\bibfnamefont {N.~Y.}\ \bibnamefont {Yao}}, \bibinfo {author} {\bibfnamefont
  {S.~D.}\ \bibnamefont {Bennett}}, \bibinfo {author} {\bibfnamefont
  {F.}~\bibnamefont {Pastawski}}, \bibinfo {author} {\bibfnamefont
  {D.}~\bibnamefont {Hunger}}, \bibinfo {author} {\bibfnamefont
  {N.}~\bibnamefont {Chisholm}}, \bibinfo {author} {\bibfnamefont
  {M.}~\bibnamefont {Markham}}, \bibinfo {author} {\bibfnamefont {D.~J.}\
  \bibnamefont {Twitchen}}, \bibinfo {author} {\bibfnamefont {J.~I.}\
  \bibnamefont {Cirac}}, \ and\ \bibinfo {author} {\bibfnamefont {M.~D.}\
  \bibnamefont {Lukin}},\ }\href@noop {} {\bibfield  {journal} {\bibinfo
  {journal} {Science}\ }\textbf {\bibinfo {volume} {336}},\ \bibinfo {pages}
  {1283} (\bibinfo {year} {2012})}\BibitemShut {NoStop}%
\bibitem [{\citenamefont {Fuchs}\ \emph {et~al.}(2011)\citenamefont {Fuchs},
  \citenamefont {Burkard}, \citenamefont {Klimov},\ and\ \citenamefont
  {Awschalom}}]{quant_memory_awschalom}%
  \BibitemOpen
  \bibfield  {author} {\bibinfo {author} {\bibfnamefont {G.~D.}\ \bibnamefont
  {Fuchs}}, \bibinfo {author} {\bibfnamefont {G.}~\bibnamefont {Burkard}},
  \bibinfo {author} {\bibfnamefont {P.~V.}\ \bibnamefont {Klimov}}, \ and\
  \bibinfo {author} {\bibfnamefont {D.~D.}\ \bibnamefont {Awschalom}},\
  }\href@noop {} {\bibfield  {journal} {\bibinfo  {journal} {Nature Physics}\
  }\textbf {\bibinfo {volume} {7}},\ \bibinfo {pages} {789} (\bibinfo {year}
  {2011})}\BibitemShut {NoStop}%
\bibitem [{\citenamefont {Jiang}\ \emph {et~al.}(2009)\citenamefont {Jiang},
  \citenamefont {Hodges}, \citenamefont {Maze}, \citenamefont {Maurer},
  \citenamefont {Taylor}, \citenamefont {Cory}, \citenamefont {Hemmer},
  \citenamefont {Walsworth}, \citenamefont {Yacoby},\ and\ \citenamefont
  {Zibrov}}]{single_shot_electron}%
  \BibitemOpen
  \bibfield  {author} {\bibinfo {author} {\bibfnamefont {L.}~\bibnamefont
  {Jiang}}, \bibinfo {author} {\bibfnamefont {J.}~\bibnamefont {Hodges}},
  \bibinfo {author} {\bibfnamefont {J.}~\bibnamefont {Maze}}, \bibinfo {author}
  {\bibfnamefont {P.}~\bibnamefont {Maurer}}, \bibinfo {author} {\bibfnamefont
  {J.}~\bibnamefont {Taylor}}, \bibinfo {author} {\bibfnamefont
  {D.}~\bibnamefont {Cory}}, \bibinfo {author} {\bibfnamefont {P.}~\bibnamefont
  {Hemmer}}, \bibinfo {author} {\bibfnamefont {R.~L.}\ \bibnamefont
  {Walsworth}}, \bibinfo {author} {\bibfnamefont {A.}~\bibnamefont {Yacoby}}, \
  and\ \bibinfo {author} {\bibfnamefont {A.~S.}\ \bibnamefont {Zibrov}},\
  }\href@noop {} {\bibfield  {journal} {\bibinfo  {journal} {Science}\ }\textbf
  {\bibinfo {volume} {326}},\ \bibinfo {pages} {267} (\bibinfo {year}
  {2009})}\BibitemShut {NoStop}%
\bibitem [{\citenamefont {Dolde}\ \emph {et~al.}(2013)\citenamefont {Dolde},
  \citenamefont {Jakobi}, \citenamefont {Naydenov}, \citenamefont {Zhao},
  \citenamefont {Pezzagna}, \citenamefont {Trautmann}, \citenamefont {Meijer},
  \citenamefont {Neumann}, \citenamefont {Jelezko},\ and\ \citenamefont
  {Wrachtrup}}]{rt_ent_2013}%
  \BibitemOpen
  \bibfield  {author} {\bibinfo {author} {\bibfnamefont {F.}~\bibnamefont
  {Dolde}}, \bibinfo {author} {\bibfnamefont {I.}~\bibnamefont {Jakobi}},
  \bibinfo {author} {\bibfnamefont {B.}~\bibnamefont {Naydenov}}, \bibinfo
  {author} {\bibfnamefont {N.}~\bibnamefont {Zhao}}, \bibinfo {author}
  {\bibfnamefont {S.}~\bibnamefont {Pezzagna}}, \bibinfo {author}
  {\bibfnamefont {C.}~\bibnamefont {Trautmann}}, \bibinfo {author}
  {\bibfnamefont {J.}~\bibnamefont {Meijer}}, \bibinfo {author} {\bibfnamefont
  {P.}~\bibnamefont {Neumann}}, \bibinfo {author} {\bibfnamefont
  {F.}~\bibnamefont {Jelezko}}, \ and\ \bibinfo {author} {\bibfnamefont
  {J.}~\bibnamefont {Wrachtrup}},\ }\href@noop {} {\bibfield  {journal}
  {\bibinfo  {journal} {Nature physics}\ }\textbf {\bibinfo {volume} {9}},\
  \bibinfo {pages} {139} (\bibinfo {year} {2013})}\BibitemShut {NoStop}%
\bibitem [{\citenamefont {Dolde}\ \emph {et~al.}(2014)\citenamefont {Dolde},
  \citenamefont {Bergholm}, \citenamefont {Wang}, \citenamefont {Jakobi},
  \citenamefont {Naydenov}, \citenamefont {Pezzagna}, \citenamefont {Meijer},
  \citenamefont {Jelezko}, \citenamefont {Neumann}, \citenamefont
  {Schulte-Herbr\"{u}ggen}, \citenamefont {Biamonte},\ and\ \citenamefont
  {Wrachtrup}}]{quant_entangle_2014}%
  \BibitemOpen
  \bibfield  {author} {\bibinfo {author} {\bibfnamefont {F.}~\bibnamefont
  {Dolde}}, \bibinfo {author} {\bibfnamefont {V.}~\bibnamefont {Bergholm}},
  \bibinfo {author} {\bibfnamefont {Y.}~\bibnamefont {Wang}}, \bibinfo {author}
  {\bibfnamefont {I.}~\bibnamefont {Jakobi}}, \bibinfo {author} {\bibfnamefont
  {B.}~\bibnamefont {Naydenov}}, \bibinfo {author} {\bibfnamefont
  {S.}~\bibnamefont {Pezzagna}}, \bibinfo {author} {\bibfnamefont
  {J.}~\bibnamefont {Meijer}}, \bibinfo {author} {\bibfnamefont
  {F.}~\bibnamefont {Jelezko}}, \bibinfo {author} {\bibfnamefont
  {P.}~\bibnamefont {Neumann}}, \bibinfo {author} {\bibfnamefont
  {T.}~\bibnamefont {Schulte-Herbr\"{u}ggen}}, \bibinfo {author} {\bibfnamefont
  {J.}~\bibnamefont {Biamonte}}, \ and\ \bibinfo {author} {\bibfnamefont
  {J.}~\bibnamefont {Wrachtrup}},\ }\href@noop {} {\bibfield  {journal}
  {\bibinfo  {journal} {Nature communications}\ }\textbf {\bibinfo {volume}
  {5}},\ \bibinfo {pages} {3371} (\bibinfo {year} {2014})}\BibitemShut
  {NoStop}%
\bibitem [{\citenamefont {Pfaff}\ \emph {et~al.}(2013)\citenamefont {Pfaff},
  \citenamefont {Taminiau}, \citenamefont {Robledo}, \citenamefont {Bernien},
  \citenamefont {Markham}, \citenamefont {Twitchen},\ and\ \citenamefont
  {Hanson}}]{hanson_ent_measurement}%
  \BibitemOpen
  \bibfield  {author} {\bibinfo {author} {\bibfnamefont {W.}~\bibnamefont
  {Pfaff}}, \bibinfo {author} {\bibfnamefont {T.~H.}\ \bibnamefont {Taminiau}},
  \bibinfo {author} {\bibfnamefont {L.}~\bibnamefont {Robledo}}, \bibinfo
  {author} {\bibfnamefont {H.}~\bibnamefont {Bernien}}, \bibinfo {author}
  {\bibfnamefont {M.}~\bibnamefont {Markham}}, \bibinfo {author} {\bibfnamefont
  {D.~J.}\ \bibnamefont {Twitchen}}, \ and\ \bibinfo {author} {\bibfnamefont
  {R.}~\bibnamefont {Hanson}},\ }\href@noop {} {\bibfield  {journal} {\bibinfo
  {journal} {Nature Physics}\ }\textbf {\bibinfo {volume} {9}},\ \bibinfo
  {pages} {29} (\bibinfo {year} {2013})}\BibitemShut {NoStop}%
\bibitem [{\citenamefont {Yang}\ \emph {et~al.}(2016)\citenamefont {Yang},
  \citenamefont {Wang}, \citenamefont {Rao}, \citenamefont {Tran},
  \citenamefont {Momenzadeh}, \citenamefont {Markham}, \citenamefont
  {Twitchen}, \citenamefont {Wang}, \citenamefont {Yang}, \citenamefont
  {Stöhr}, \citenamefont {Neumann}, \citenamefont {Kosaka},\ and\
  \citenamefont {Wrachtrup}}]{ys_qrepeater}%
  \BibitemOpen
  \bibfield  {author} {\bibinfo {author} {\bibfnamefont {S.}~\bibnamefont
  {Yang}}, \bibinfo {author} {\bibfnamefont {Y.}~\bibnamefont {Wang}}, \bibinfo
  {author} {\bibfnamefont {D.~D.~B.}\ \bibnamefont {Rao}}, \bibinfo {author}
  {\bibfnamefont {T.~H.}\ \bibnamefont {Tran}}, \bibinfo {author}
  {\bibfnamefont {A.~S.}\ \bibnamefont {Momenzadeh}}, \bibinfo {author}
  {\bibfnamefont {M.}~\bibnamefont {Markham}}, \bibinfo {author} {\bibfnamefont
  {D.~J.}\ \bibnamefont {Twitchen}}, \bibinfo {author} {\bibfnamefont
  {P.}~\bibnamefont {Wang}}, \bibinfo {author} {\bibfnamefont {W.}~\bibnamefont
  {Yang}}, \bibinfo {author} {\bibfnamefont {R.}~\bibnamefont {Stöhr}},
  \bibinfo {author} {\bibfnamefont {P.}~\bibnamefont {Neumann}}, \bibinfo
  {author} {\bibfnamefont {H.}~\bibnamefont {Kosaka}}, \ and\ \bibinfo {author}
  {\bibfnamefont {J.}~\bibnamefont {Wrachtrup}},\ }\href@noop {} {\bibfield
  {journal} {\bibinfo  {journal} {Nature Photonics}\ }\textbf {\bibinfo
  {volume} {10}},\ \bibinfo {pages} {507} (\bibinfo {year} {2016})}\BibitemShut
  {NoStop}%
\bibitem [{\citenamefont {Kalb}\ \emph {et~al.}(2017)\citenamefont {Kalb},
  \citenamefont {Reiserer}, \citenamefont {Humphreys}, \citenamefont
  {Bakermans}, \citenamefont {Kamerling}, \citenamefont {Nickerson},
  \citenamefont {Benjamin}, \citenamefont {Twitchen}, \citenamefont {Markham},\
  and\ \citenamefont {Hanson}}]{hanson_quantnetwork}%
  \BibitemOpen
  \bibfield  {author} {\bibinfo {author} {\bibfnamefont {N.}~\bibnamefont
  {Kalb}}, \bibinfo {author} {\bibfnamefont {A.~A.}\ \bibnamefont {Reiserer}},
  \bibinfo {author} {\bibfnamefont {P.~C.}\ \bibnamefont {Humphreys}}, \bibinfo
  {author} {\bibfnamefont {J.~J.~W.}\ \bibnamefont {Bakermans}}, \bibinfo
  {author} {\bibfnamefont {S.~J.}\ \bibnamefont {Kamerling}}, \bibinfo {author}
  {\bibfnamefont {N.~H.}\ \bibnamefont {Nickerson}}, \bibinfo {author}
  {\bibfnamefont {S.~C.}\ \bibnamefont {Benjamin}}, \bibinfo {author}
  {\bibfnamefont {D.~J.}\ \bibnamefont {Twitchen}}, \bibinfo {author}
  {\bibfnamefont {M.}~\bibnamefont {Markham}}, \ and\ \bibinfo {author}
  {\bibfnamefont {R.}~\bibnamefont {Hanson}},\ }\href@noop {} {\bibfield
  {journal} {\bibinfo  {journal} {Science}\ }\textbf {\bibinfo {volume}
  {356}},\ \bibinfo {pages} {928} (\bibinfo {year} {2017})}\BibitemShut
  {NoStop}%
\bibitem [{\citenamefont {Dutt}\ \emph {et~al.}(2007)\citenamefont {Dutt},
  \citenamefont {Childress}, \citenamefont {Jiang}, \citenamefont {Togan},
  \citenamefont {Maze}, \citenamefont {Jelezko}, \citenamefont {Zibrov},
  \citenamefont {Hemmer},\ and\ \citenamefont {Lukin}}]{lukin_science_2007}%
  \BibitemOpen
  \bibfield  {author} {\bibinfo {author} {\bibfnamefont {M.~V.~G.}\
  \bibnamefont {Dutt}}, \bibinfo {author} {\bibfnamefont {L.}~\bibnamefont
  {Childress}}, \bibinfo {author} {\bibfnamefont {L.}~\bibnamefont {Jiang}},
  \bibinfo {author} {\bibfnamefont {E.}~\bibnamefont {Togan}}, \bibinfo
  {author} {\bibfnamefont {J.}~\bibnamefont {Maze}}, \bibinfo {author}
  {\bibfnamefont {F.}~\bibnamefont {Jelezko}}, \bibinfo {author} {\bibfnamefont
  {A.~S.}\ \bibnamefont {Zibrov}}, \bibinfo {author} {\bibfnamefont {P.~R.}\
  \bibnamefont {Hemmer}}, \ and\ \bibinfo {author} {\bibfnamefont {M.~D.}\
  \bibnamefont {Lukin}},\ }\href@noop {} {\bibfield  {journal} {\bibinfo
  {journal} {Science}\ }\textbf {\bibinfo {volume} {316}},\ \bibinfo {pages}
  {1312} (\bibinfo {year} {2007})}\BibitemShut {NoStop}%
\bibitem [{\citenamefont {Neumann}\ \emph {et~al.}(2008)\citenamefont
  {Neumann}, \citenamefont {Mizuochi}, \citenamefont {Rempp}, \citenamefont
  {Hemmer}, \citenamefont {Watanabe}, \citenamefont {Yamasaki}, \citenamefont
  {Jacques}, \citenamefont {Gaebel}, \citenamefont {Jelezko},\ and\
  \citenamefont {Wrachtrup}}]{nv_3bit_entangle}%
  \BibitemOpen
  \bibfield  {author} {\bibinfo {author} {\bibfnamefont {P.}~\bibnamefont
  {Neumann}}, \bibinfo {author} {\bibfnamefont {N.}~\bibnamefont {Mizuochi}},
  \bibinfo {author} {\bibfnamefont {F.}~\bibnamefont {Rempp}}, \bibinfo
  {author} {\bibfnamefont {P.}~\bibnamefont {Hemmer}}, \bibinfo {author}
  {\bibfnamefont {H.}~\bibnamefont {Watanabe}}, \bibinfo {author}
  {\bibfnamefont {S.}~\bibnamefont {Yamasaki}}, \bibinfo {author}
  {\bibfnamefont {V.}~\bibnamefont {Jacques}}, \bibinfo {author} {\bibfnamefont
  {T.}~\bibnamefont {Gaebel}}, \bibinfo {author} {\bibfnamefont
  {F.}~\bibnamefont {Jelezko}}, \ and\ \bibinfo {author} {\bibfnamefont
  {J.}~\bibnamefont {Wrachtrup}},\ }\href@noop {} {\bibfield  {journal}
  {\bibinfo  {journal} {Science}\ }\textbf {\bibinfo {volume} {320}},\ \bibinfo
  {pages} {1326} (\bibinfo {year} {2008})}\BibitemShut {NoStop}%
\bibitem [{\citenamefont {Waldherr}\ \emph {et~al.}(2014)\citenamefont
  {Waldherr}, \citenamefont {Wang}, \citenamefont {Zaiser}, \citenamefont
  {Jamali}, \citenamefont {Schulte-Herbr\"{u}ggen}, \citenamefont {Abe},
  \citenamefont {Ohshima}, \citenamefont {Isoya}, \citenamefont {Du},
  \citenamefont {Neumann},\ and\ \citenamefont {Wrachtrup}}]{nv_qec}%
  \BibitemOpen
  \bibfield  {author} {\bibinfo {author} {\bibfnamefont {G.}~\bibnamefont
  {Waldherr}}, \bibinfo {author} {\bibfnamefont {Y.}~\bibnamefont {Wang}},
  \bibinfo {author} {\bibfnamefont {S.}~\bibnamefont {Zaiser}}, \bibinfo
  {author} {\bibfnamefont {M.}~\bibnamefont {Jamali}}, \bibinfo {author}
  {\bibfnamefont {T.}~\bibnamefont {Schulte-Herbr\"{u}ggen}}, \bibinfo {author}
  {\bibfnamefont {H.}~\bibnamefont {Abe}}, \bibinfo {author} {\bibfnamefont
  {T.}~\bibnamefont {Ohshima}}, \bibinfo {author} {\bibfnamefont
  {J.}~\bibnamefont {Isoya}}, \bibinfo {author} {\bibfnamefont {J.~F.}\
  \bibnamefont {Du}}, \bibinfo {author} {\bibfnamefont {P.}~\bibnamefont
  {Neumann}}, \ and\ \bibinfo {author} {\bibfnamefont {J.}~\bibnamefont
  {Wrachtrup}},\ }\href@noop {} {\bibfield  {journal} {\bibinfo  {journal}
  {Nature}\ }\textbf {\bibinfo {volume} {506}},\ \bibinfo {pages} {204}
  (\bibinfo {year} {2014})}\BibitemShut {NoStop}%
\bibitem [{\citenamefont {Taminiau}\ \emph {et~al.}(2014)\citenamefont
  {Taminiau}, \citenamefont {Cramer}, \citenamefont {van~der Sar},
  \citenamefont {Dobrovitski},\ and\ \citenamefont {Hanson}}]{hanson_qec}%
  \BibitemOpen
  \bibfield  {author} {\bibinfo {author} {\bibfnamefont {T.~H.}\ \bibnamefont
  {Taminiau}}, \bibinfo {author} {\bibfnamefont {J.}~\bibnamefont {Cramer}},
  \bibinfo {author} {\bibfnamefont {T.}~\bibnamefont {van~der Sar}}, \bibinfo
  {author} {\bibfnamefont {V.~V.}\ \bibnamefont {Dobrovitski}}, \ and\ \bibinfo
  {author} {\bibfnamefont {R.}~\bibnamefont {Hanson}},\ }\href@noop {}
  {\bibfield  {journal} {\bibinfo  {journal} {Nature nanotechnology}\ }\textbf
  {\bibinfo {volume} {9}},\ \bibinfo {pages} {171} (\bibinfo {year}
  {2014})}\BibitemShut {NoStop}%
\bibitem [{\citenamefont {Taminiau}\ \emph {et~al.}(2012)\citenamefont
  {Taminiau}, \citenamefont {Wagenaar}, \citenamefont {van~der Sar},
  \citenamefont {Jelezko}, \citenamefont {Dobrovitski},\ and\ \citenamefont
  {Hanson}}]{hanson_weak_nuclear_spin}%
  \BibitemOpen
  \bibfield  {author} {\bibinfo {author} {\bibfnamefont {T.~H.}\ \bibnamefont
  {Taminiau}}, \bibinfo {author} {\bibfnamefont {J.~J.~T.}\ \bibnamefont
  {Wagenaar}}, \bibinfo {author} {\bibfnamefont {T.}~\bibnamefont {van~der
  Sar}}, \bibinfo {author} {\bibfnamefont {F.}~\bibnamefont {Jelezko}},
  \bibinfo {author} {\bibfnamefont {V.~V.}\ \bibnamefont {Dobrovitski}}, \ and\
  \bibinfo {author} {\bibfnamefont {R.}~\bibnamefont {Hanson}},\ }\href@noop {}
  {\bibfield  {journal} {\bibinfo  {journal} {Physical review letters}\
  }\textbf {\bibinfo {volume} {109}},\ \bibinfo {pages} {137602} (\bibinfo
  {year} {2012})}\BibitemShut {NoStop}%
\bibitem [{\citenamefont {Kong}\ \emph {et~al.}(2016)\citenamefont {Kong},
  \citenamefont {Ju}, \citenamefont {Liu}, \citenamefont {Lei}, \citenamefont
  {Wang}, \citenamefont {Kong}, \citenamefont {Wang}, \citenamefont {Huang},
  \citenamefont {Li}, \citenamefont {Shi}, \citenamefont {Jiang},\ and\
  \citenamefont {Du}}]{kf_prl_2016}%
  \BibitemOpen
  \bibfield  {author} {\bibinfo {author} {\bibfnamefont {F.}~\bibnamefont
  {Kong}}, \bibinfo {author} {\bibfnamefont {C.}~\bibnamefont {Ju}}, \bibinfo
  {author} {\bibfnamefont {Y.}~\bibnamefont {Liu}}, \bibinfo {author}
  {\bibfnamefont {C.}~\bibnamefont {Lei}}, \bibinfo {author} {\bibfnamefont
  {M.}~\bibnamefont {Wang}}, \bibinfo {author} {\bibfnamefont {X.}~\bibnamefont
  {Kong}}, \bibinfo {author} {\bibfnamefont {P.}~\bibnamefont {Wang}}, \bibinfo
  {author} {\bibfnamefont {P.}~\bibnamefont {Huang}}, \bibinfo {author}
  {\bibfnamefont {Z.}~\bibnamefont {Li}}, \bibinfo {author} {\bibfnamefont
  {F.}~\bibnamefont {Shi}}, \bibinfo {author} {\bibfnamefont {L.}~\bibnamefont
  {Jiang}}, \ and\ \bibinfo {author} {\bibfnamefont {J.}~\bibnamefont {Du}},\
  }\href@noop {} {\bibfield  {journal} {\bibinfo  {journal} {Physical review
  letters}\ }\textbf {\bibinfo {volume} {117}},\ \bibinfo {pages} {060503}
  (\bibinfo {year} {2016})}\BibitemShut {NoStop}%
\bibitem [{\citenamefont {Xu}\ \emph {et~al.}(2017)\citenamefont {Xu},
  \citenamefont {Xie}, \citenamefont {Li}, \citenamefont {Xu}, \citenamefont
  {Wang}, \citenamefont {Ye}, \citenamefont {Kong}, \citenamefont {Geng},
  \citenamefont {Duan}, \citenamefont {Shi},\ and\ \citenamefont
  {Du}}]{nv_aqc}%
  \BibitemOpen
  \bibfield  {author} {\bibinfo {author} {\bibfnamefont {K.}~\bibnamefont
  {Xu}}, \bibinfo {author} {\bibfnamefont {T.}~\bibnamefont {Xie}}, \bibinfo
  {author} {\bibfnamefont {Z.}~\bibnamefont {Li}}, \bibinfo {author}
  {\bibfnamefont {X.}~\bibnamefont {Xu}}, \bibinfo {author} {\bibfnamefont
  {M.}~\bibnamefont {Wang}}, \bibinfo {author} {\bibfnamefont {X.}~\bibnamefont
  {Ye}}, \bibinfo {author} {\bibfnamefont {F.}~\bibnamefont {Kong}}, \bibinfo
  {author} {\bibfnamefont {J.}~\bibnamefont {Geng}}, \bibinfo {author}
  {\bibfnamefont {C.}~\bibnamefont {Duan}}, \bibinfo {author} {\bibfnamefont
  {F.}~\bibnamefont {Shi}}, \ and\ \bibinfo {author} {\bibfnamefont
  {J.}~\bibnamefont {Du}},\ }\href@noop {} {\bibfield  {journal} {\bibinfo
  {journal} {Physical review letters}\ }\textbf {\bibinfo {volume} {118}},\
  \bibinfo {pages} {130504} (\bibinfo {year} {2017})}\BibitemShut {NoStop}%
\bibitem [{\citenamefont {Zaiser}\ \emph {et~al.}(2016)\citenamefont {Zaiser},
  \citenamefont {Rendler}, \citenamefont {Jakobi}, \citenamefont {Wolf},
  \citenamefont {Lee}, \citenamefont {Wagner}, \citenamefont {Bergholm},
  \citenamefont {Schulte-Herbr\"{u}ggen}, \citenamefont {Neumann},\ and\
  \citenamefont {Wrachtrup}}]{Joreg_ent_sensing}%
  \BibitemOpen
  \bibfield  {author} {\bibinfo {author} {\bibfnamefont {S.}~\bibnamefont
  {Zaiser}}, \bibinfo {author} {\bibfnamefont {T.}~\bibnamefont {Rendler}},
  \bibinfo {author} {\bibfnamefont {I.}~\bibnamefont {Jakobi}}, \bibinfo
  {author} {\bibfnamefont {T.}~\bibnamefont {Wolf}}, \bibinfo {author}
  {\bibfnamefont {S.-Y.}\ \bibnamefont {Lee}}, \bibinfo {author} {\bibfnamefont
  {S.}~\bibnamefont {Wagner}}, \bibinfo {author} {\bibfnamefont
  {V.}~\bibnamefont {Bergholm}}, \bibinfo {author} {\bibfnamefont
  {T.}~\bibnamefont {Schulte-Herbr\"{u}ggen}}, \bibinfo {author} {\bibfnamefont
  {P.}~\bibnamefont {Neumann}}, \ and\ \bibinfo {author} {\bibfnamefont
  {J.}~\bibnamefont {Wrachtrup}},\ }\href@noop {} {\bibfield  {journal}
  {\bibinfo  {journal} {Nature communications}\ }\textbf {\bibinfo {volume}
  {7}},\ \bibinfo {pages} {12279} (\bibinfo {year} {2016})}\BibitemShut
  {NoStop}%
\bibitem [{\citenamefont {Jakobi}\ \emph {et~al.}(2017)\citenamefont {Jakobi},
  \citenamefont {Neumann}, \citenamefont {Wang}, \citenamefont {Dasari},
  \citenamefont {El~Hallak}, \citenamefont {Bashir}, \citenamefont {Markham},
  \citenamefont {Edmonds}, \citenamefont {Twitchen},\ and\ \citenamefont
  {Wrachtrup}}]{Joerg_broadband}%
  \BibitemOpen
  \bibfield  {author} {\bibinfo {author} {\bibfnamefont {I.}~\bibnamefont
  {Jakobi}}, \bibinfo {author} {\bibfnamefont {P.}~\bibnamefont {Neumann}},
  \bibinfo {author} {\bibfnamefont {Y.}~\bibnamefont {Wang}}, \bibinfo {author}
  {\bibfnamefont {D.~B.~R.}\ \bibnamefont {Dasari}}, \bibinfo {author}
  {\bibfnamefont {F.}~\bibnamefont {El~Hallak}}, \bibinfo {author}
  {\bibfnamefont {M.~A.}\ \bibnamefont {Bashir}}, \bibinfo {author}
  {\bibfnamefont {M.}~\bibnamefont {Markham}}, \bibinfo {author} {\bibfnamefont
  {A.}~\bibnamefont {Edmonds}}, \bibinfo {author} {\bibfnamefont
  {D.}~\bibnamefont {Twitchen}}, \ and\ \bibinfo {author} {\bibfnamefont
  {J.}~\bibnamefont {Wrachtrup}},\ }\href@noop {} {\bibfield  {journal}
  {\bibinfo  {journal} {Nature nanotechnology}\ }\textbf {\bibinfo {volume}
  {12}},\ \bibinfo {pages} {67} (\bibinfo {year} {2017})}\BibitemShut {NoStop}%
\bibitem [{\citenamefont {Unden}\ \emph {et~al.}(2016)\citenamefont {Unden},
  \citenamefont {Balasubramanian}, \citenamefont {Louzon}, \citenamefont
  {Vinkler}, \citenamefont {Plenio}, \citenamefont {Markham}, \citenamefont
  {Twitchen}, \citenamefont {Stacey}, \citenamefont {Lovchinsky}, \citenamefont
  {Sushkov}, \citenamefont {Lukin}, \citenamefont {Retzker}, \citenamefont
  {Naydenov}, \citenamefont {McGuinness},\ and\ \citenamefont
  {Jelezko}}]{fador_qec_sensing}%
  \BibitemOpen
  \bibfield  {author} {\bibinfo {author} {\bibfnamefont {T.}~\bibnamefont
  {Unden}}, \bibinfo {author} {\bibfnamefont {P.}~\bibnamefont
  {Balasubramanian}}, \bibinfo {author} {\bibfnamefont {D.}~\bibnamefont
  {Louzon}}, \bibinfo {author} {\bibfnamefont {Y.}~\bibnamefont {Vinkler}},
  \bibinfo {author} {\bibfnamefont {M.}~\bibnamefont {Plenio}}, \bibinfo
  {author} {\bibfnamefont {M.}~\bibnamefont {Markham}}, \bibinfo {author}
  {\bibfnamefont {D.}~\bibnamefont {Twitchen}}, \bibinfo {author}
  {\bibfnamefont {A.}~\bibnamefont {Stacey}}, \bibinfo {author} {\bibfnamefont
  {I.}~\bibnamefont {Lovchinsky}}, \bibinfo {author} {\bibfnamefont
  {A.}~\bibnamefont {Sushkov}}, \bibinfo {author} {\bibfnamefont
  {M.}~\bibnamefont {Lukin}}, \bibinfo {author} {\bibfnamefont
  {A.}~\bibnamefont {Retzker}}, \bibinfo {author} {\bibfnamefont
  {B.}~\bibnamefont {Naydenov}}, \bibinfo {author} {\bibfnamefont
  {L.}~\bibnamefont {McGuinness}}, \ and\ \bibinfo {author} {\bibfnamefont
  {F.}~\bibnamefont {Jelezko}},\ }\href@noop {} {\bibfield  {journal} {\bibinfo
   {journal} {Physical Review Letters}\ }\textbf {\bibinfo {volume} {116}},\
  \bibinfo {pages} {230502} (\bibinfo {year} {2016})}\BibitemShut {NoStop}%
\bibitem [{\citenamefont {Pfender}\ \emph {et~al.}(2017)\citenamefont
  {Pfender}, \citenamefont {Aslam}, \citenamefont {Sumiya}, \citenamefont
  {Onoda}, \citenamefont {Neumann}, \citenamefont {Isoya}, \citenamefont
  {Meriles},\ and\ \citenamefont {Wrachtrup}}]{Joerg_sensing_cluster}%
  \BibitemOpen
  \bibfield  {author} {\bibinfo {author} {\bibfnamefont {M.}~\bibnamefont
  {Pfender}}, \bibinfo {author} {\bibfnamefont {N.}~\bibnamefont {Aslam}},
  \bibinfo {author} {\bibfnamefont {H.}~\bibnamefont {Sumiya}}, \bibinfo
  {author} {\bibfnamefont {S.}~\bibnamefont {Onoda}}, \bibinfo {author}
  {\bibfnamefont {P.}~\bibnamefont {Neumann}}, \bibinfo {author} {\bibfnamefont
  {J.}~\bibnamefont {Isoya}}, \bibinfo {author} {\bibfnamefont {C.~A.}\
  \bibnamefont {Meriles}}, \ and\ \bibinfo {author} {\bibfnamefont
  {J.}~\bibnamefont {Wrachtrup}},\ }\href@noop {} {\bibfield  {journal}
  {\bibinfo  {journal} {Nature Communications}\ }\textbf {\bibinfo {volume}
  {8}},\ \bibinfo {pages} {834} (\bibinfo {year} {2017})}\BibitemShut {NoStop}%
\bibitem [{\citenamefont {Rosskopf}\ \emph {et~al.}(2017)\citenamefont
  {Rosskopf}, \citenamefont {Zopes}, \citenamefont {Boss},\ and\ \citenamefont
  {Degen}}]{Degen_npj_2017}%
  \BibitemOpen
  \bibfield  {author} {\bibinfo {author} {\bibfnamefont {T.}~\bibnamefont
  {Rosskopf}}, \bibinfo {author} {\bibfnamefont {J.}~\bibnamefont {Zopes}},
  \bibinfo {author} {\bibfnamefont {J.~M.}\ \bibnamefont {Boss}}, \ and\
  \bibinfo {author} {\bibfnamefont {C.~L.}\ \bibnamefont {Degen}},\ }\href@noop
  {} {\bibfield  {journal} {\bibinfo  {journal} {npj Quantum Information}\
  }\textbf {\bibinfo {volume} {3}},\ \bibinfo {pages} {33} (\bibinfo {year}
  {2017})}\BibitemShut {NoStop}%
\bibitem [{\citenamefont {Aslam}\ \emph {et~al.}(2017)\citenamefont {Aslam},
  \citenamefont {Pfender}, \citenamefont {Neumann}, \citenamefont {Reuter},
  \citenamefont {Zappe}, \citenamefont {de~Oliveira}, \citenamefont
  {Denisenko}, \citenamefont {Sumiya}, \citenamefont {Onoda}, \citenamefont
  {Isoya},\ and\ \citenamefont {Wrachtrup}}]{Joerg_chemical_shift}%
  \BibitemOpen
  \bibfield  {author} {\bibinfo {author} {\bibfnamefont {N.}~\bibnamefont
  {Aslam}}, \bibinfo {author} {\bibfnamefont {M.}~\bibnamefont {Pfender}},
  \bibinfo {author} {\bibfnamefont {P.}~\bibnamefont {Neumann}}, \bibinfo
  {author} {\bibfnamefont {R.}~\bibnamefont {Reuter}}, \bibinfo {author}
  {\bibfnamefont {A.}~\bibnamefont {Zappe}}, \bibinfo {author} {\bibfnamefont
  {F.~F.}\ \bibnamefont {de~Oliveira}}, \bibinfo {author} {\bibfnamefont
  {A.}~\bibnamefont {Denisenko}}, \bibinfo {author} {\bibfnamefont
  {H.}~\bibnamefont {Sumiya}}, \bibinfo {author} {\bibfnamefont
  {S.}~\bibnamefont {Onoda}}, \bibinfo {author} {\bibfnamefont
  {J.}~\bibnamefont {Isoya}}, \ and\ \bibinfo {author} {\bibfnamefont
  {J.}~\bibnamefont {Wrachtrup}},\ }\href@noop {} {\bibfield  {journal}
  {\bibinfo  {journal} {Science}\ }\textbf {\bibinfo {volume} {357}},\ \bibinfo
  {pages} {67} (\bibinfo {year} {2017})}\BibitemShut {NoStop}%
\bibitem [{\citenamefont {Gruber}\ \emph {et~al.}(1997)\citenamefont {Gruber},
  \citenamefont {Drabenstedt}, \citenamefont {Tietz}, \citenamefont {Fleury},
  \citenamefont {Wrachtrup},\ and\ \citenamefont
  {vonBorczyskowski}}]{Joerg_science_1997}%
  \BibitemOpen
  \bibfield  {author} {\bibinfo {author} {\bibfnamefont {A.}~\bibnamefont
  {Gruber}}, \bibinfo {author} {\bibfnamefont {A.}~\bibnamefont {Drabenstedt}},
  \bibinfo {author} {\bibfnamefont {C.}~\bibnamefont {Tietz}}, \bibinfo
  {author} {\bibfnamefont {L.}~\bibnamefont {Fleury}}, \bibinfo {author}
  {\bibfnamefont {J.}~\bibnamefont {Wrachtrup}}, \ and\ \bibinfo {author}
  {\bibfnamefont {C.}~\bibnamefont {vonBorczyskowski}},\ }\href@noop {}
  {\bibfield  {journal} {\bibinfo  {journal} {Science}\ }\textbf {\bibinfo
  {volume} {276}},\ \bibinfo {pages} {2012} (\bibinfo {year}
  {1997})}\BibitemShut {NoStop}%
\bibitem [{\citenamefont {Chakraborty}\ \emph {et~al.}(2017)\citenamefont
  {Chakraborty}, \citenamefont {Zhang},\ and\ \citenamefont
  {Suter}}]{suter_npol}%
  \BibitemOpen
  \bibfield  {author} {\bibinfo {author} {\bibfnamefont {T.}~\bibnamefont
  {Chakraborty}}, \bibinfo {author} {\bibfnamefont {J.}~\bibnamefont {Zhang}},
  \ and\ \bibinfo {author} {\bibfnamefont {D.}~\bibnamefont {Suter}},\
  }\href@noop {} {\bibfield  {journal} {\bibinfo  {journal} {New Journal of
  Physics}\ }\textbf {\bibinfo {volume} {19}},\ \bibinfo {pages} {073030}
  (\bibinfo {year} {2017})}\BibitemShut {NoStop}%
\bibitem [{\citenamefont {\'{A}lvarez}\ \emph {et~al.}(2015)\citenamefont
  {\'{A}lvarez}, \citenamefont {Bretschneider}, \citenamefont {Fischer},
  \citenamefont {London}, \citenamefont {Kanda}, \citenamefont {Onoda},
  \citenamefont {Isoya}, \citenamefont {Gershoni},\ and\ \citenamefont
  {Frydman}}]{nc_frydman_2015}%
  \BibitemOpen
  \bibfield  {author} {\bibinfo {author} {\bibfnamefont {G.~A.}\ \bibnamefont
  {\'{A}lvarez}}, \bibinfo {author} {\bibfnamefont {C.~O.}\ \bibnamefont
  {Bretschneider}}, \bibinfo {author} {\bibfnamefont {R.}~\bibnamefont
  {Fischer}}, \bibinfo {author} {\bibfnamefont {P.}~\bibnamefont {London}},
  \bibinfo {author} {\bibfnamefont {H.}~\bibnamefont {Kanda}}, \bibinfo
  {author} {\bibfnamefont {S.}~\bibnamefont {Onoda}}, \bibinfo {author}
  {\bibfnamefont {J.}~\bibnamefont {Isoya}}, \bibinfo {author} {\bibfnamefont
  {D.}~\bibnamefont {Gershoni}}, \ and\ \bibinfo {author} {\bibfnamefont
  {L.}~\bibnamefont {Frydman}},\ }\href@noop {} {\bibfield  {journal} {\bibinfo
   {journal} {Nature Communications}\ }\textbf {\bibinfo {volume} {6}},\
  \bibinfo {pages} {8456} (\bibinfo {year} {2015})}\BibitemShut {NoStop}%
\bibitem [{\citenamefont {King}\ \emph {et~al.}(2015)\citenamefont {King},
  \citenamefont {Jeong}, \citenamefont {Vassiliou}, \citenamefont {Shin},
  \citenamefont {Page}, \citenamefont {Avalos}, \citenamefont {Wang},\ and\
  \citenamefont {Pines}}]{nc_pines_2015}%
  \BibitemOpen
  \bibfield  {author} {\bibinfo {author} {\bibfnamefont {J.~P.}\ \bibnamefont
  {King}}, \bibinfo {author} {\bibfnamefont {K.}~\bibnamefont {Jeong}},
  \bibinfo {author} {\bibfnamefont {C.~C.}\ \bibnamefont {Vassiliou}}, \bibinfo
  {author} {\bibfnamefont {C.~S.}\ \bibnamefont {Shin}}, \bibinfo {author}
  {\bibfnamefont {R.~H.}\ \bibnamefont {Page}}, \bibinfo {author}
  {\bibfnamefont {C.~E.}\ \bibnamefont {Avalos}}, \bibinfo {author}
  {\bibfnamefont {H.-J.}\ \bibnamefont {Wang}}, \ and\ \bibinfo {author}
  {\bibfnamefont {A.}~\bibnamefont {Pines}},\ }\href@noop {} {\bibfield
  {journal} {\bibinfo  {journal} {Nature Communications}\ }\textbf {\bibinfo
  {volume} {6}},\ \bibinfo {pages} {8965} (\bibinfo {year} {2015})}\BibitemShut
  {NoStop}%
\bibitem [{\citenamefont {Broadway}\ \emph {et~al.}(2018)\citenamefont
  {Broadway}, \citenamefont {Tetienne}, \citenamefont {Stacey}, \citenamefont
  {Wood}, \citenamefont {Simpson}, \citenamefont {Hall},\ and\ \citenamefont
  {Hollenberg}}]{nc_hollenberg_2018}%
  \BibitemOpen
  \bibfield  {author} {\bibinfo {author} {\bibfnamefont {D.~A.}\ \bibnamefont
  {Broadway}}, \bibinfo {author} {\bibfnamefont {J.-P.}\ \bibnamefont
  {Tetienne}}, \bibinfo {author} {\bibfnamefont {A.}~\bibnamefont {Stacey}},
  \bibinfo {author} {\bibfnamefont {J.~D.~A.}\ \bibnamefont {Wood}}, \bibinfo
  {author} {\bibfnamefont {D.~A.}\ \bibnamefont {Simpson}}, \bibinfo {author}
  {\bibfnamefont {L.~T.}\ \bibnamefont {Hall}}, \ and\ \bibinfo {author}
  {\bibfnamefont {L.~C.~L.}\ \bibnamefont {Hollenberg}},\ }\href@noop {}
  {\bibfield  {journal} {\bibinfo  {journal} {Nature Communications}\ }\textbf
  {\bibinfo {volume} {9}},\ \bibinfo {pages} {1246} (\bibinfo {year}
  {2018})}\BibitemShut {NoStop}%
\bibitem [{\citenamefont {Dreau}\ \emph {et~al.}(2012)\citenamefont {Dreau},
  \citenamefont {Maze}, \citenamefont {Lesik}, \citenamefont {Roch},\ and\
  \citenamefont {Jacques}}]{prb_nuclear_spin}%
  \BibitemOpen
  \bibfield  {author} {\bibinfo {author} {\bibfnamefont {A.}~\bibnamefont
  {Dreau}}, \bibinfo {author} {\bibfnamefont {J.~R.}\ \bibnamefont {Maze}},
  \bibinfo {author} {\bibfnamefont {M.}~\bibnamefont {Lesik}}, \bibinfo
  {author} {\bibfnamefont {J.~F.}\ \bibnamefont {Roch}}, \ and\ \bibinfo
  {author} {\bibfnamefont {V.}~\bibnamefont {Jacques}},\ }\href@noop {}
  {\bibfield  {journal} {\bibinfo  {journal} {Physical Review B}\ }\textbf
  {\bibinfo {volume} {85}},\ \bibinfo {pages} {134107} (\bibinfo {year}
  {2012})}\BibitemShut {NoStop}%
\bibitem [{\citenamefont {Abobeih}\ \emph {et~al.}(2018)\citenamefont
  {Abobeih}, \citenamefont {Cramer}, \citenamefont {Bakker}, \citenamefont
  {Kalb}, \citenamefont {Twitchen}, \citenamefont {Markham},\ and\
  \citenamefont {Taminiau}}]{Taminiau_22_nuclspin}%
  \BibitemOpen
  \bibfield  {author} {\bibinfo {author} {\bibfnamefont {M.~H.}\ \bibnamefont
  {Abobeih}}, \bibinfo {author} {\bibfnamefont {J.}~\bibnamefont {Cramer}},
  \bibinfo {author} {\bibfnamefont {M.~A.}\ \bibnamefont {Bakker}}, \bibinfo
  {author} {\bibfnamefont {N.}~\bibnamefont {Kalb}}, \bibinfo {author}
  {\bibfnamefont {D.~J.}\ \bibnamefont {Twitchen}}, \bibinfo {author}
  {\bibfnamefont {M.}~\bibnamefont {Markham}}, \ and\ \bibinfo {author}
  {\bibfnamefont {T.~H.}\ \bibnamefont {Taminiau}},\ }\href@noop {} {\bibfield
  {journal} {\bibinfo  {journal} {arXiv preprint arXiv:1801.01196}\ } (\bibinfo
  {year} {2018})}\BibitemShut {NoStop}%
\bibitem [{\citenamefont {Jacques}\ \emph {et~al.}(2009)\citenamefont
  {Jacques}, \citenamefont {Neumann}, \citenamefont {Beck}, \citenamefont
  {Markham}, \citenamefont {Twitchen}, \citenamefont {Meijer}, \citenamefont
  {Kaiser}, \citenamefont {Balasubramanian}, \citenamefont {Jelezko},\ and\
  \citenamefont {Wrachtrup}}]{nv_n_pol}%
  \BibitemOpen
  \bibfield  {author} {\bibinfo {author} {\bibfnamefont {V.}~\bibnamefont
  {Jacques}}, \bibinfo {author} {\bibfnamefont {P.}~\bibnamefont {Neumann}},
  \bibinfo {author} {\bibfnamefont {J.}~\bibnamefont {Beck}}, \bibinfo {author}
  {\bibfnamefont {M.}~\bibnamefont {Markham}}, \bibinfo {author} {\bibfnamefont
  {D.}~\bibnamefont {Twitchen}}, \bibinfo {author} {\bibfnamefont
  {J.}~\bibnamefont {Meijer}}, \bibinfo {author} {\bibfnamefont
  {F.}~\bibnamefont {Kaiser}}, \bibinfo {author} {\bibfnamefont
  {G.}~\bibnamefont {Balasubramanian}}, \bibinfo {author} {\bibfnamefont
  {F.}~\bibnamefont {Jelezko}}, \ and\ \bibinfo {author} {\bibfnamefont
  {J.}~\bibnamefont {Wrachtrup}},\ }\href@noop {} {\bibfield  {journal}
  {\bibinfo  {journal} {Physical Review Letters}\ }\textbf {\bibinfo {volume}
  {102}},\ \bibinfo {pages} {057403} (\bibinfo {year} {2009})}\BibitemShut
  {NoStop}%
\bibitem [{\citenamefont {Hanson}\ \emph {et~al.}(2006)\citenamefont {Hanson},
  \citenamefont {Mendoza}, \citenamefont {Epstein},\ and\ \citenamefont
  {Awschalom}}]{Awschalom_polarization}%
  \BibitemOpen
  \bibfield  {author} {\bibinfo {author} {\bibfnamefont {R.}~\bibnamefont
  {Hanson}}, \bibinfo {author} {\bibfnamefont {F.}~\bibnamefont {Mendoza}},
  \bibinfo {author} {\bibfnamefont {R.}~\bibnamefont {Epstein}}, \ and\
  \bibinfo {author} {\bibfnamefont {D.}~\bibnamefont {Awschalom}},\ }\href@noop
  {} {\bibfield  {journal} {\bibinfo  {journal} {Physical review letters}\
  }\textbf {\bibinfo {volume} {97}},\ \bibinfo {pages} {087601} (\bibinfo
  {year} {2006})}\BibitemShut {NoStop}%
\bibitem [{\citenamefont {Fischer}\ \emph {et~al.}(2013)\citenamefont
  {Fischer}, \citenamefont {Bretschneider}, \citenamefont {London},
  \citenamefont {Budker}, \citenamefont {Gershoni},\ and\ \citenamefont
  {Frydman}}]{bulk_c_pol2013}%
  \BibitemOpen
  \bibfield  {author} {\bibinfo {author} {\bibfnamefont {R.}~\bibnamefont
  {Fischer}}, \bibinfo {author} {\bibfnamefont {C.~O.}\ \bibnamefont
  {Bretschneider}}, \bibinfo {author} {\bibfnamefont {P.}~\bibnamefont
  {London}}, \bibinfo {author} {\bibfnamefont {D.}~\bibnamefont {Budker}},
  \bibinfo {author} {\bibfnamefont {D.}~\bibnamefont {Gershoni}}, \ and\
  \bibinfo {author} {\bibfnamefont {L.}~\bibnamefont {Frydman}},\ }\href@noop
  {} {\bibfield  {journal} {\bibinfo  {journal} {Physical Review Letters}\
  }\textbf {\bibinfo {volume} {111}},\ \bibinfo {pages} {057601} (\bibinfo
  {year} {2013})}\BibitemShut {NoStop}%
\bibitem [{\citenamefont {Wang}\ \emph
  {et~al.}(2015{\natexlab{a}})\citenamefont {Wang}, \citenamefont {Liu},\ and\
  \citenamefont {Yang}}]{yangwen_scientic_report}%
  \BibitemOpen
  \bibfield  {author} {\bibinfo {author} {\bibfnamefont {P.}~\bibnamefont
  {Wang}}, \bibinfo {author} {\bibfnamefont {B.}~\bibnamefont {Liu}}, \ and\
  \bibinfo {author} {\bibfnamefont {W.}~\bibnamefont {Yang}},\ }\href@noop {}
  {\bibfield  {journal} {\bibinfo  {journal} {Scientific Reports}\ }\textbf
  {\bibinfo {volume} {5}},\ \bibinfo {pages} {15847} (\bibinfo {year}
  {2015}{\natexlab{a}})}\BibitemShut {NoStop}%
\bibitem [{\citenamefont {Wang}\ \emph {et~al.}(2013)\citenamefont {Wang},
  \citenamefont {Shin}, \citenamefont {Avalos}, \citenamefont {Seltzer},
  \citenamefont {Budker}, \citenamefont {Pines},\ and\ \citenamefont
  {Bajaj}}]{ensem_c_pol2013}%
  \BibitemOpen
  \bibfield  {author} {\bibinfo {author} {\bibfnamefont {H.-J.}\ \bibnamefont
  {Wang}}, \bibinfo {author} {\bibfnamefont {C.~S.}\ \bibnamefont {Shin}},
  \bibinfo {author} {\bibfnamefont {C.~E.}\ \bibnamefont {Avalos}}, \bibinfo
  {author} {\bibfnamefont {S.~J.}\ \bibnamefont {Seltzer}}, \bibinfo {author}
  {\bibfnamefont {D.}~\bibnamefont {Budker}}, \bibinfo {author} {\bibfnamefont
  {A.}~\bibnamefont {Pines}}, \ and\ \bibinfo {author} {\bibfnamefont {V.~S.}\
  \bibnamefont {Bajaj}},\ }\href@noop {} {\bibfield  {journal} {\bibinfo
  {journal} {Nature Communications}\ }\textbf {\bibinfo {volume} {4}},\
  \bibinfo {pages} {1940} (\bibinfo {year} {2013})}\BibitemShut {NoStop}%
\bibitem [{\citenamefont {Geng}\ \emph {et~al.}(2016)\citenamefont {Geng},
  \citenamefont {Wu}, \citenamefont {Wang}, \citenamefont {Xu}, \citenamefont
  {Shi}, \citenamefont {Xie}, \citenamefont {Rong},\ and\ \citenamefont
  {Du}}]{gjp_prl_2016}%
  \BibitemOpen
  \bibfield  {author} {\bibinfo {author} {\bibfnamefont {J.}~\bibnamefont
  {Geng}}, \bibinfo {author} {\bibfnamefont {Y.}~\bibnamefont {Wu}}, \bibinfo
  {author} {\bibfnamefont {X.}~\bibnamefont {Wang}}, \bibinfo {author}
  {\bibfnamefont {K.}~\bibnamefont {Xu}}, \bibinfo {author} {\bibfnamefont
  {F.}~\bibnamefont {Shi}}, \bibinfo {author} {\bibfnamefont {Y.}~\bibnamefont
  {Xie}}, \bibinfo {author} {\bibfnamefont {X.}~\bibnamefont {Rong}}, \ and\
  \bibinfo {author} {\bibfnamefont {J.}~\bibnamefont {Du}},\ }\href@noop {}
  {\bibfield  {journal} {\bibinfo  {journal} {Physical review letters}\
  }\textbf {\bibinfo {volume} {117}},\ \bibinfo {pages} {170501} (\bibinfo
  {year} {2016})}\BibitemShut {NoStop}%
\bibitem [{\citenamefont {Rong}\ \emph {et~al.}(2015)\citenamefont {Rong},
  \citenamefont {Geng}, \citenamefont {Shi}, \citenamefont {Liu}, \citenamefont
  {Xu}, \citenamefont {Ma}, \citenamefont {Kong}, \citenamefont {Jiang},
  \citenamefont {Wu},\ and\ \citenamefont {Du}}]{rx_nc_2015}%
  \BibitemOpen
  \bibfield  {author} {\bibinfo {author} {\bibfnamefont {X.}~\bibnamefont
  {Rong}}, \bibinfo {author} {\bibfnamefont {J.}~\bibnamefont {Geng}}, \bibinfo
  {author} {\bibfnamefont {F.}~\bibnamefont {Shi}}, \bibinfo {author}
  {\bibfnamefont {Y.}~\bibnamefont {Liu}}, \bibinfo {author} {\bibfnamefont
  {K.}~\bibnamefont {Xu}}, \bibinfo {author} {\bibfnamefont {W.}~\bibnamefont
  {Ma}}, \bibinfo {author} {\bibfnamefont {F.}~\bibnamefont {Kong}}, \bibinfo
  {author} {\bibfnamefont {Z.}~\bibnamefont {Jiang}}, \bibinfo {author}
  {\bibfnamefont {Y.}~\bibnamefont {Wu}}, \ and\ \bibinfo {author}
  {\bibfnamefont {J.}~\bibnamefont {Du}},\ }\href@noop {} {\bibfield  {journal}
  {\bibinfo  {journal} {Nature communications}\ }\textbf {\bibinfo {volume}
  {6}},\ \bibinfo {pages} {8748} (\bibinfo {year} {2015})}\BibitemShut
  {NoStop}%
\bibitem [{\citenamefont {Zu}\ \emph {et~al.}(2014)\citenamefont {Zu},
  \citenamefont {Wang}, \citenamefont {He}, \citenamefont {Zhang},
  \citenamefont {Dai}, \citenamefont {Wang},\ and\ \citenamefont
  {Duan}}]{duan_geometric_gates}%
  \BibitemOpen
  \bibfield  {author} {\bibinfo {author} {\bibfnamefont {C.}~\bibnamefont
  {Zu}}, \bibinfo {author} {\bibfnamefont {W.-B.}\ \bibnamefont {Wang}},
  \bibinfo {author} {\bibfnamefont {L.}~\bibnamefont {He}}, \bibinfo {author}
  {\bibfnamefont {W.-G.}\ \bibnamefont {Zhang}}, \bibinfo {author}
  {\bibfnamefont {C.-Y.}\ \bibnamefont {Dai}}, \bibinfo {author} {\bibfnamefont
  {F.}~\bibnamefont {Wang}}, \ and\ \bibinfo {author} {\bibfnamefont {L.-M.}\
  \bibnamefont {Duan}},\ }\href@noop {} {\bibfield  {journal} {\bibinfo
  {journal} {Nature}\ }\textbf {\bibinfo {volume} {514}},\ \bibinfo {pages}
  {72} (\bibinfo {year} {2014})}\BibitemShut {NoStop}%
\bibitem [{\citenamefont {Jiang}\ \emph {et~al.}(2008)\citenamefont {Jiang},
  \citenamefont {Dutt}, \citenamefont {Togan}, \citenamefont {Childress},
  \citenamefont {Cappellaro}, \citenamefont {Taylor},\ and\ \citenamefont
  {Lukin}}]{Jiang_nuclear_coherence}%
  \BibitemOpen
  \bibfield  {author} {\bibinfo {author} {\bibfnamefont {L.}~\bibnamefont
  {Jiang}}, \bibinfo {author} {\bibfnamefont {M.~G.}\ \bibnamefont {Dutt}},
  \bibinfo {author} {\bibfnamefont {E.}~\bibnamefont {Togan}}, \bibinfo
  {author} {\bibfnamefont {L.}~\bibnamefont {Childress}}, \bibinfo {author}
  {\bibfnamefont {P.}~\bibnamefont {Cappellaro}}, \bibinfo {author}
  {\bibfnamefont {J.~M.}\ \bibnamefont {Taylor}}, \ and\ \bibinfo {author}
  {\bibfnamefont {M.~D.}\ \bibnamefont {Lukin}},\ }\href@noop {} {\bibfield
  {journal} {\bibinfo  {journal} {Physical Review Letters}\ }\textbf {\bibinfo
  {volume} {100}},\ \bibinfo {pages} {073001} (\bibinfo {year}
  {2008})}\BibitemShut {NoStop}%
\bibitem [{\citenamefont {Neumann}\ \emph {et~al.}(2010)\citenamefont
  {Neumann}, \citenamefont {Beck}, \citenamefont {Steiner}, \citenamefont
  {Rempp}, \citenamefont {Fedder}, \citenamefont {Hemmer}, \citenamefont
  {Wrachtrup},\ and\ \citenamefont {Jelezko}}]{single_shot_nuclear}%
  \BibitemOpen
  \bibfield  {author} {\bibinfo {author} {\bibfnamefont {P.}~\bibnamefont
  {Neumann}}, \bibinfo {author} {\bibfnamefont {J.}~\bibnamefont {Beck}},
  \bibinfo {author} {\bibfnamefont {M.}~\bibnamefont {Steiner}}, \bibinfo
  {author} {\bibfnamefont {F.}~\bibnamefont {Rempp}}, \bibinfo {author}
  {\bibfnamefont {H.}~\bibnamefont {Fedder}}, \bibinfo {author} {\bibfnamefont
  {P.~R.}\ \bibnamefont {Hemmer}}, \bibinfo {author} {\bibfnamefont
  {J.}~\bibnamefont {Wrachtrup}}, \ and\ \bibinfo {author} {\bibfnamefont
  {F.}~\bibnamefont {Jelezko}},\ }\href@noop {} {\bibfield  {journal} {\bibinfo
   {journal} {Science}\ }\textbf {\bibinfo {volume} {329}},\ \bibinfo {pages}
  {542} (\bibinfo {year} {2010})}\BibitemShut {NoStop}%
\bibitem [{\citenamefont {Laraoui}\ \emph {et~al.}(2013)\citenamefont
  {Laraoui}, \citenamefont {Dolde}, \citenamefont {Burk}, \citenamefont
  {Reinhard}, \citenamefont {Wrachtrup},\ and\ \citenamefont
  {Meriles}}]{cspin_corrspec}%
  \BibitemOpen
  \bibfield  {author} {\bibinfo {author} {\bibfnamefont {A.}~\bibnamefont
  {Laraoui}}, \bibinfo {author} {\bibfnamefont {F.}~\bibnamefont {Dolde}},
  \bibinfo {author} {\bibfnamefont {C.}~\bibnamefont {Burk}}, \bibinfo {author}
  {\bibfnamefont {F.}~\bibnamefont {Reinhard}}, \bibinfo {author}
  {\bibfnamefont {J.}~\bibnamefont {Wrachtrup}}, \ and\ \bibinfo {author}
  {\bibfnamefont {C.~A.}\ \bibnamefont {Meriles}},\ }\href@noop {} {\bibfield
  {journal} {\bibinfo  {journal} {Nature communications}\ }\textbf {\bibinfo
  {volume} {4}},\ \bibinfo {pages} {1651} (\bibinfo {year} {2013})}\BibitemShut
  {NoStop}%
\bibitem [{\citenamefont {Pagliero}\ \emph {et~al.}(2014)\citenamefont
  {Pagliero}, \citenamefont {Laraoui}, \citenamefont {Henshaw},\ and\
  \citenamefont {Meriles}}]{dr_n_pol}%
  \BibitemOpen
  \bibfield  {author} {\bibinfo {author} {\bibfnamefont {D.}~\bibnamefont
  {Pagliero}}, \bibinfo {author} {\bibfnamefont {A.}~\bibnamefont {Laraoui}},
  \bibinfo {author} {\bibfnamefont {J.~D.}\ \bibnamefont {Henshaw}}, \ and\
  \bibinfo {author} {\bibfnamefont {C.~A.}\ \bibnamefont {Meriles}},\
  }\href@noop {} {\bibfield  {journal} {\bibinfo  {journal} {Applied Physics
  Letters}\ }\textbf {\bibinfo {volume} {105}},\ \bibinfo {pages} {242402}
  (\bibinfo {year} {2014})}\BibitemShut {NoStop}%
\bibitem [{\citenamefont {Wang}\ \emph
  {et~al.}(2015{\natexlab{b}})\citenamefont {Wang}, \citenamefont {Dolde},
  \citenamefont {Biamonte}, \citenamefont {Babbush}, \citenamefont {Bergholm},
  \citenamefont {Yang}, \citenamefont {Jakobi}, \citenamefont {Neumann},
  \citenamefont {Aspuru-Guzik}, \citenamefont {Whitfield},\ and\ \citenamefont
  {Wrachtrup}}]{wy_qsim_2015}%
  \BibitemOpen
  \bibfield  {author} {\bibinfo {author} {\bibfnamefont {Y.}~\bibnamefont
  {Wang}}, \bibinfo {author} {\bibfnamefont {F.}~\bibnamefont {Dolde}},
  \bibinfo {author} {\bibfnamefont {J.}~\bibnamefont {Biamonte}}, \bibinfo
  {author} {\bibfnamefont {R.}~\bibnamefont {Babbush}}, \bibinfo {author}
  {\bibfnamefont {V.}~\bibnamefont {Bergholm}}, \bibinfo {author}
  {\bibfnamefont {S.}~\bibnamefont {Yang}}, \bibinfo {author} {\bibfnamefont
  {I.}~\bibnamefont {Jakobi}}, \bibinfo {author} {\bibfnamefont
  {P.}~\bibnamefont {Neumann}}, \bibinfo {author} {\bibfnamefont
  {A.}~\bibnamefont {Aspuru-Guzik}}, \bibinfo {author} {\bibfnamefont {J.~D.}\
  \bibnamefont {Whitfield}}, \ and\ \bibinfo {author} {\bibfnamefont
  {J.}~\bibnamefont {Wrachtrup}},\ }\href@noop {} {\bibfield  {journal}
  {\bibinfo  {journal} {ACS nano}\ }\textbf {\bibinfo {volume} {9}},\ \bibinfo
  {pages} {7769} (\bibinfo {year} {2015}{\natexlab{b}})}\BibitemShut {NoStop}%
\bibitem [{\citenamefont {Manson}\ and\ \citenamefont
  {Harrison}(2005)}]{Manson_nv0}%
  \BibitemOpen
  \bibfield  {author} {\bibinfo {author} {\bibfnamefont {N.}~\bibnamefont
  {Manson}}\ and\ \bibinfo {author} {\bibfnamefont {J.}~\bibnamefont
  {Harrison}},\ }\href@noop {} {\bibfield  {journal} {\bibinfo  {journal}
  {Diamond and related materials}\ }\textbf {\bibinfo {volume} {14}},\ \bibinfo
  {pages} {1705} (\bibinfo {year} {2005})}\BibitemShut {NoStop}%
\bibitem [{\citenamefont {Waldherr}\ \emph {et~al.}(2011)\citenamefont
  {Waldherr}, \citenamefont {Beck}, \citenamefont {Steiner}, \citenamefont
  {Neumann}, \citenamefont {Gali}, \citenamefont {Frauenheim}, \citenamefont
  {Jelezko},\ and\ \citenamefont {Wrachtrup}}]{Waldherr_nv0}%
  \BibitemOpen
  \bibfield  {author} {\bibinfo {author} {\bibfnamefont {G.}~\bibnamefont
  {Waldherr}}, \bibinfo {author} {\bibfnamefont {J.}~\bibnamefont {Beck}},
  \bibinfo {author} {\bibfnamefont {M.}~\bibnamefont {Steiner}}, \bibinfo
  {author} {\bibfnamefont {P.}~\bibnamefont {Neumann}}, \bibinfo {author}
  {\bibfnamefont {A.}~\bibnamefont {Gali}}, \bibinfo {author} {\bibfnamefont
  {T.}~\bibnamefont {Frauenheim}}, \bibinfo {author} {\bibfnamefont
  {F.}~\bibnamefont {Jelezko}}, \ and\ \bibinfo {author} {\bibfnamefont
  {J.}~\bibnamefont {Wrachtrup}},\ }\href@noop {} {\bibfield  {journal}
  {\bibinfo  {journal} {Physical Review Letters}\ }\textbf {\bibinfo {volume}
  {106}},\ \bibinfo {pages} {157601} (\bibinfo {year} {2011})}\BibitemShut
  {NoStop}%
\bibitem [{\citenamefont {Aslam}\ \emph {et~al.}(2013)\citenamefont {Aslam},
  \citenamefont {Waldherr}, \citenamefont {Neumann}, \citenamefont {Jelezko},\
  and\ \citenamefont {Wrachtrup}}]{Joerg_nv0}%
  \BibitemOpen
  \bibfield  {author} {\bibinfo {author} {\bibfnamefont {N.}~\bibnamefont
  {Aslam}}, \bibinfo {author} {\bibfnamefont {G.}~\bibnamefont {Waldherr}},
  \bibinfo {author} {\bibfnamefont {P.}~\bibnamefont {Neumann}}, \bibinfo
  {author} {\bibfnamefont {F.}~\bibnamefont {Jelezko}}, \ and\ \bibinfo
  {author} {\bibfnamefont {J.}~\bibnamefont {Wrachtrup}},\ }\href@noop {}
  {\bibfield  {journal} {\bibinfo  {journal} {New Journal of Physics}\ }\textbf
  {\bibinfo {volume} {15}},\ \bibinfo {pages} {013064} (\bibinfo {year}
  {2013})}\BibitemShut {NoStop}%
\bibitem [{\citenamefont {Siyushev}\ \emph {et~al.}(2013)\citenamefont
  {Siyushev}, \citenamefont {Pinto}, \citenamefont {V\"{o}r\"{o}s},
  \citenamefont {Gali}, \citenamefont {Jelezko},\ and\ \citenamefont
  {Wrachtrup}}]{Siyushev_nv0}%
  \BibitemOpen
  \bibfield  {author} {\bibinfo {author} {\bibfnamefont {P.}~\bibnamefont
  {Siyushev}}, \bibinfo {author} {\bibfnamefont {H.}~\bibnamefont {Pinto}},
  \bibinfo {author} {\bibfnamefont {M.}~\bibnamefont {V\"{o}r\"{o}s}}, \bibinfo
  {author} {\bibfnamefont {A.}~\bibnamefont {Gali}}, \bibinfo {author}
  {\bibfnamefont {F.}~\bibnamefont {Jelezko}}, \ and\ \bibinfo {author}
  {\bibfnamefont {J.}~\bibnamefont {Wrachtrup}},\ }\href@noop {} {\bibfield
  {journal} {\bibinfo  {journal} {Physical Review Letters}\ }\textbf {\bibinfo
  {volume} {110}},\ \bibinfo {pages} {167402} (\bibinfo {year}
  {2013})}\BibitemShut {NoStop}%
\bibitem [{si_()}]{si_npol}%
  \BibitemOpen
  \href@noop {} {\bibinfo  {journal} {See Supplemental Online Material for
  theoretical model of the illuminated nuclear depolarization, numerical
  simulation and experimental details of the chopped laser pulses.}\
  }\BibitemShut {NoStop}%
\bibitem [{\citenamefont {Hunter}(2007)}]{matplotlib}%
  \BibitemOpen
\bibfield  {journal} {  }\bibfield  {author} {\bibinfo {author} {\bibfnamefont
  {J.~D.}\ \bibnamefont {Hunter}},\ }\href@noop {} {\bibfield  {journal}
  {\bibinfo  {journal} {Computing in science and engineering}\ }\textbf
  {\bibinfo {volume} {9}},\ \bibinfo {pages} {90} (\bibinfo {year}
  {2007})}\BibitemShut {NoStop}%
\end{thebibliography}

%

\end{document}